\documentclass[aps,prb,twocolumn,groupedaddress,showpacs]{revtex4}

\usepackage[bookmarks]{hyperref}
\usepackage{graphicx}
\usepackage{amsmath}
\usepackage{amstext}
\usepackage{graphics}
\usepackage{exscale}
\usepackage{epsfig}
\usepackage[T1]{fontenc}
\usepackage{bm}
\usepackage{bbm}

\usepackage{version}

\usepackage{latexsym}

\usepackage[tight]{subfigure}

\renewcommand{\epsilon}{\varepsilon}

\newcommand{\gfbraket}[1]{\langle\!\langle #1  \rangle\!\rangle}

\newcommand{\unitop}{\mathbbm{1}}

\newcommand{\integral}[3]{\!\int\limits_{#2}^{#3}\!\!{\rm d}#1\;}

\newcommand{\expval}[2]{ \langle  #1 #2\ \!\! \rangle}

\newcommand{\elcre}[2]{ c^{\dagger}_{#1,#2}}
\newcommand{\elann}[2]{ c_{#1,#2}}

\newcommand{\e}{\mathrm e}

\newcommand{\vct}[1]{\bm #1}
\newcommand{\vk}{{\bm k}}
\newcommand{\vq}{{\bm q}}

\newcommand{\thGf}{{\cal G}}

\newcommand{\Imag}{\mathrm{Im}}

\newcommand{\hc}{\mathrm{h.c.}}

\includeversion{commentst1} 

\begin{document}

\title{Retardation effects and the Coulomb pseudopotential in the theory of superconductivity}  
\author{Johannes Bauer,${}^{1,2}$ Jong E. Han,${}^{1,3}$ and Olle Gunnarsson${}^1$}
\affiliation{${}^1$Max Planck Institute for Solid State Research, Heisenbergstr.1,
  D-70569 Stuttgart, Germany} 
\affiliation{${}^2$Department of Physics, Harvard University, Cambridge,
  Massachusetts 02138, USA}
\affiliation{${}^3$  Department of Physics, SUNY at Buffalo, Buffalo,
New York 14260, USA}
\date{\today} 
\begin{abstract}
In the theory of electron-phonon superconductivity both the magnitude of the
electron-phonon coupling $\lambda$ as well as the Coulomb pseudopotential
$\mu^*$ are important to determine the transition temperature $T_c$ and other
properties. We calculate corrections to the conventional result for the Coulomb
pseudopotential. Our calculation are based on the Hubbard-Holstein model,
where electron-electron and electron-phonon interactions are local. We develop
a perturbation expansion, which 
accounts for the important renormalization effects for the
electrons, the phonons, and the electron-phonon vertex. 
We show that retardation effects are still operative for higher
order corrections, but less efficient due to a reduction of the effective
bandwidth. This can lead to larger values of the pseudopotential and reduced
values of $T_c$. The conclusions from the perturbative calculations 
are corroborated up to intermediate couplings by comparison with
non-perturbative dynamical mean-field results.     
\end{abstract}
\pacs{71.10.Fd, 71.27.+a,71.30.+h,75.20.-g, 71.10.Ay}

\maketitle

\section{Introduction}

More than a century after its discovery superconductivity continues to
be subject of intense research in condensed matter physics. The
research ranges from topics geared to the technical application
of the phenomenon over the search for new superconducting materials to
fundamental questions of microscopic mechanisms.\cite{Nor11} The latter are
important as one might hope that a good understanding of the mechanisms
will make the design of new superconductors at elevated temperatures 
feasible.\cite{Gur11,Can11} There are numerous classes of materials for which
different mechanisms are discussed. \cite{BK08,Nor11} In most cases a
bosonic pairing function is invoked, which could be of purely electronic
origin such as spin fluctuations, or the conventional phonon
mechanism, but also more exotic mechanisms have been proposed.
The electron-phonon mechanism has the longest history and probably the
most established  mathematical foundation.  An effective
electron-electron attraction generated by the electron-phonon
interaction is part of the celebrated Bardeen-Cooper-Schrieffer (BCS)
theory.\cite{BCS57} 
The more elaborate theory including microscopic details goes under the
name Migdal-Eliashberg (ME) theory. \cite{Eli60} With the help of ME theory the
superconducting properties of many elements and numerous alloys have been
described accurately.\cite{Car90,MC08} 

The general ideas of ME theory can be presented in relatively simple
fashion, although the details of the complete framework, its
foundations and specific applications, involve a substantial degree of
sophistication to which many researcher have contributed over the
years.\cite{schrieffer,AM82,Car90,MC08} One cornerstone is 
Migdal's theorem,\cite{Mig58} which employs the fact that the typical electronic
energy scale $E_{\rm el}$ and the typical phonon energy scale
$\omega_{\rm  ph}$ differ largely, such that $E_{\rm el}/\omega_{\rm
  ph}$ is of the order 100 and larger - the electrons move much faster
than the  phonons.  It can then be shown that the perturbation theory of the
electron-phonon problem greatly simplifies since 
vertex corrections are small. The influence of the
bosonic pairing function  $\alpha^2F(\omega)$
on the electronic properties and the occurrence of superconductivity
can therefore be computed  reliably. A remarkable aspect of
the ME theory is that it is even well justified for large values of
the electron-phonon coupling parameter $\lambda>1$ as long as the effective
expansion parameter $\sim \lambda \omega_{\rm   ph}/E_{\rm el}$
remains small.\cite{Mig58,MK82,BHG11} With the help of the ME equations, the
pairing function $\alpha^2F(\omega)$  
and the phenomenological parameter for the Coulomb pseudopotential $\mu^*$
were extracted from tunneling 
measurement in a procedure termed inverse tunneling
spectroscopy.\cite{MR69} Based on those other properties such as $T_c$ or
thermodynamic quantities can be computed. The agreement with the
respective experimental values of many elements and alloys, notably Pb
and Nb, is within the range of a few percent.\cite{Car90} This leads to a consistent
picture and was taken as proof of the validity of ME theory and the electron-phonon
mechanism.\cite{Car90,MC08} Further support comes from first principle
calculations for  $\alpha^2F(\omega)$, which are in many cases in
remarkable agreement with the results extracted from
tunneling.\cite{Car90,SSA94,SS96,MC08} 

In addition to the effective attraction induced by the electron-phonon
coupling the applications of ME theory include the effects of the
Coulomb repulsion. Due to the enormous success of the theory it is sometimes
understated that in contrast to the electron-phonon problem no rigorous
arguments exist for the treatment of the Coulomb repulsion,\cite{AM82} the reason
being the absence of a small parameter as in Migdal's
theorem. Traditionally, this is seen as minor deficiency based on the
following arguments. Some effects of the Coulomb interaction, such as
the renormalization of the electron-phonon couplings, $g$, the
electron and phonon dispersion $\epsilon_{\vk},\omega_{\vq}$ are 
implicit in the experimentally derived results or in the first
principle calculations. It remains to deal with the direct repulsion in
the pairing channel which opposes s-wave superconductivity. 
Morel and Anderson proposed a procedure in two stages:\cite{MA62} first the
Coulomb interaction is screened and averaged over the Fermi surface. In a
second step it is projected to the phonon scale. The result possesses the
famous form,\cite{BTS62,MA62,schrieffer}   
\begin{equation}
  \mu_c^*=\frac{\mu_c}{1+\mu_c\log\Big(\frac{E_{\rm el}}{\omega_{\rm
        ph}}\Big)},
\label{eq:mustar}
\end{equation}
which is sometimes termed the Morel-Anderson (MA) pseudopotential.
Here, $\mu_c$ is a dimensionless quantity which consists of a product of the
the averaged, screened Coulomb interaction with the density of states per spin
at the Fermi energy; $E_{\rm el}$ is an electronic
scale, such as the half bandwidth $D$ or the Fermi energy $E_{\rm F}$, and
$\omega_{\rm ph}$ the phonon scale, e.g. the Debye frequency. 
Eq. (\ref{eq:mustar}) has the important property that for the typical large energy
separation between electronic and phonon scale, $E_{\rm el} \gg \omega_{\rm
  ph}$, such that $\log(E_{\rm el}/\omega_{\rm ph})\sim 5-10$,
$\mu_c^*$ is much smaller than $\mu_c$. This effect is 
remarkable, as it enables the electron-phonon s-wave superconductivity
to be possible in spite of the Coulomb repulsion, 
which on a bare level is much larger and working directly against it. 
This is sometimes even termed the true mechanism of electron-phonon
superconductivity.\cite{MC08}
The textbook physical picture is that electrons do not need to be
close in position space and suffer from the Coulomb repulsion in order
to pair since the electron-phonon 
interaction is retarded and thus electrons can pair with a
``time-delay''.  
Using appropriate energy scales in Eq. (\ref{eq:mustar}) leads to estimates of the order
$\mu_c^*\sim 0.1-0.14$. This fits well to the parameters $\mu^*$
obtained from tunneling in many elemental superconductors and alloys.\cite{Car90}
Hence, the Coulomb effects are generally considered to be well
described by $\mu^*$, which is a fairly universal quantity. In
contrast to the pairing function it is usually not calculated based on
first principles and rather used as a fitting parameter. It is worth
mentioning that there is also an alternative ab-initio approach to 
superconductivity based on the DFT framework.\cite{Lea05,Mea05,Pea06a}

Review of the literature however also reveals evidence that the
description of the Coulomb repulsion in 
the form of Eq.~(\ref{eq:mustar}) is incomplete. In some cases, such
as V or Nb${}_3$Ge,\cite{Car90}  the values for $\mu^*$ appear to be
of the order $0.2-0.3$ substantially 
larger than the traditional quotes, even though the maximal phonon scales does
not seem to be particularly large.
Density functional theory (DFT) calculation\cite{SS96} for $\alpha^2F(\omega)$
find good agreement with the tunneling 
results, but to explain the experimental values for $T_c$ also
somewhat larger values of $\mu^*$ have to be used. A specific case
which raises doubts about the conventional framework is the example
of elemental lithium at ambient pressure.\cite{LC91,RA97} The coupling
constant was estimated to be 
$\lambda\sim 0.4$,\cite{LC91,BNC11} which seems to be in line with specific heat
measurements. With the usual value of $\mu^*\sim 0.1$ and the
appropriate phonon scale this implies $T_c\sim$~1K. This is in
contrast to experiments where for a long time no superconductivity was
observed down to values of 6mK,\cite{Tea70} and only recently, $T_c\sim$~0.4mK
was found at ambient pressure, which requires $\mu^*\sim 0.23$.\cite{TJUPS07}
In this context, we also mention alkali doped picene, which was
recently discovered to be superconducting at $T_c\sim$~7K and
$T_c\sim$~18K depending on preparation.\cite{Mea10} First principle calculations
only seem to be able to explain these values of $T_c$ based on an electron-phonon
mechanism with relatively large values of $\mu^*\sim 0.23$.\cite{SB11}
However, also other interpretations exist.\cite{GC11,CCPM11} 


It is our endeavor to reanalyze the expression for the
Coulomb pseudopotential in Eq.~(\ref{eq:mustar}) and calculate corrections to
it.  We focus on the reduction of phonon induced s-wave
superconductivity due to the Coulomb repulsion between
electrons. Superconductivity which is induced in an anisotropic higher order
angular momentum channel by purely repulsive interactions, such as the well-known
Kohn-Luttinger effect \cite{KL65}, is not dealt with in the present work.    
Berk and Schrieffer \cite{BS66} included a specific
class of higher order diagrams describing the coupling to ferromagnetic (FM) spin
fluctuations, addressing almost FM metals, like Pd. They   
found that retardation is ineffective for the added diagrams 
and that superconductivity is strongly suppressed, which can help to explain
the cases when FM spin fluctuations are important. 
In another line of work going beyond MA the jellium model for
electrons was considered and the Eliashberg equation were solved including 
the frequency and momentum dependence of the dynamically screened Coulomb
interaction. Curiously, random phase approximation (RPA) calculations by
Rietschel and Sham yielded negative values for $\mu^*_c$ when the density
parameter $r_s$ exceeded values $r_s>2.5$.\cite{RS83} The resulting
unrealistic large values for $T_c$ were interpreted as plasmon induced
superconductivity.\cite{GS84} It was shown 
that vertex corrections lead to a reduction of $T_c$ and partly cure this
problem.\cite{GS84,GS88E} Phenomenological models of the 
form of Kukkonen and Overhauser\cite{KO79} for the screened Coulomb
interaction mostly yield realistic positive results for $\mu^*_c\sim 0.1$ and
no s-wave superconductivity,\cite{BR90,RA97} although there also exist
different conclusions.\cite{Tak93}  
We remark that Richardson and Ashcroft\cite{RA97} arrived at a rather
accurate prediction for $T_c$ in Li at ambient pressure based on such
calculations. Nevertheless, there does not seem to be a conclusive study,
which systematically analyzes higher order corrections to the unphysical RPA
result.

Rather than treating the Coulomb interaction in the electron gas, our approach
is based on a model with local interactions.\cite{BHG12} It can thus be 
interpreted as only considering the second step of the MA procedure, where the
projection of the screened interaction to the phonon scale is considered.  
To be specific we take the Hubbard-Holstein (HH) model, where
there is a local electron-electron repulsion and an electron-phonon
interaction present. The advantage of the restriction to this model is that it
can be analyzed by means of the dynamical mean
field theory (DMFT).\cite{GKKR96} Since the DMFT becomes exact in the limit of
large dimensions, it can provide benchmark results
independent of the interaction strength and thus allows us to test otherwise
uncontrolled perturbative approximations. Retardation effects encoded in the
frequency dependence of the 
self-energies are fully contained. DMFT treats both the electron-phonon and
the electron-electron interaction non-perturbatively and therefore
include contributions to $\mu_c^*$ to all orders. However, it also includes
renormalization effects of the phonons, the electrons and the electron-phonon
coupling. This makes the interpretation of the results and the 
extraction of $\mu_c^*$ more difficult. In order to be able to
nevertheless obtain meaningful insights, we developed a diagrammatic perturbative
approach for the HH model in the limit of large dimensions. This
allows us to see how accurate the conventional theory describes the
benchmark results and which corrections are necessary to get a good
agreement between the perturbative results and DMFT. We will see that
higher order corrections to $\mu_c^*$ enter in a modified form when
compared to Eq. (\ref{eq:mustar}). 

The occurrence of superconductivity in the HH model has been analyzed
theoretically beyond ME theory.\cite{FJ95,BVL95,Tak96,KT06,TTCC07}  Freericks and
Jarrel\cite{FJ95} studied the suppression of the instabilities towards 
charge density wave (CDW) formation and superconductivity at and away from
half filling 
within the framework of DMFT. Their finding of robustness of CDW against
superconductivity could be explained within weak coupling perturbation theory
without invoking corrections to the pseudopotential result in
Eq.~(\ref{eq:mustar}).   
Functional renormalization group studies \cite{KT06,TTCC07} found spin density
wave (SDW), CDW and different superconducting instabilities. The phonon scale
in those works was however relatively large such that retardation effects only
played a moderate role.

In this work we show that retardation effects lead to the reduction $\mu_c\to
\mu_c^{\ast}$  also in the higher order calculation, but not as efficiently
as for the first order one. Non-perturbative DMFT calculations clarify
that the perturbative result is accurate up to intermediate coupling
strength. An important conclusion is then that retardation effects indeed lead
to rather small values of $\mu_c^*$, even when contributions beyond the
standard theory are considered. For systems with sizable Coulomb interactions
$\mu_c$, our values for $\mu_c^*$ are however larger than in the standard
theory and therefore lead to reduced values of the superconducting gap and
$T_c$.   
The paper is structured as follows: In Sec. II, the details of the
model are introduced as well as some basic properties of the DMFT
approach. In Sec. III, we discuss the diagrammatics for the calculation
of $T_c$ from the pairing equation, including self-energy and vertex
corrections. In Sec. IV, similar diagrams are discussed for the
calculation of the gap at $T=0$. In Sec. V, we focus on the calculation
of the $\mu^*_c$ and derive analytic results including the higher order
corrections. In Sec. VI, we put together the results from the
perturbation theory and validate our findings up to intermediate
couplings with the non-perturbative DMFT results, followed by the
conclusions.

\section{Model and DMFT setup}
The purpose of our work is to obtain generic insights into the
behavior of a coupled electron-phonon system. Specifically, we employ the
Hubbard-Holstein model, 
\begin{eqnarray}
  \label{hubholham}
  H&=&-\sum_{i,j,{\sigma}}(t_{ij}\elcre i{\sigma}\elann
j{\sigma}+\hc)+U\sum_i\hat n_{i,\uparrow}\hat n_{i,\downarrow} \\
&&+\omega_0\sum_ib_i^{\dagger}b_i+g\sum_i(b_i+b_i^{\dagger})\Big(\sum_{\sigma}\hat
n_{i,\sigma}-1\Big).
\nonumber
\end{eqnarray}
$\elcre i{\sigma}$ creates an electron at lattice site $i$ with spin $\sigma$,
and $b_i^{\dagger}$ a phonon with oscillator frequency $\omega_0$,
$\hat n_{i,\sigma}=\elcre i{\sigma}\elann 
i{\sigma}$. The electrons interact locally with strength $U$, and
their density is coupled to an optical phonon mode with coupling constant
$g$. The local oscillator displacement is related to the bosonic operators by  $\hat
x_i=(b_i+b_i^{\dagger})/\sqrt{2\omega_0}$, where $\hbar=1$, and one can
define a characteristic length $x_0=1/\sqrt{\omega_0}$ for the oscillator. 
We have set the ionic mass to $M=1$ in (\ref{hubholham}).
The model in Eq. (\ref{hubholham}) possesses the minimal ingredients
necessary, such as energy scales for 
electrons, phonons and their interactions. In the limit of large dimensions
this model can be solved exactly by DMFT. Hence, we can provide controlled benchmark
results in this situation. The following calculations are based on this model
in the limit of large dimension. In this case the self-energy is independent
of the momentum $\vk$, but retains the full frequency dependence. This can be
compared with the usual application of ME theory, where one usually projects
to the Fermi surface and only deals with frequency dependent quantities.

\subsection{Calculating $T_c$}
In DMFT the critical temperature can be calculated by analyzing the relevant
susceptibility. For completeness we display some of the results, which we use
later. The notation follows the one in Ref.~\onlinecite{GKKR96}. 
The equation for the s-wave superconductivity susceptibility
$\chi_{\vq}(i\omega_n)$ reads with
$\chi_{\vq}(i\omega_n)=\sum_{n_1,n_2}\tilde\chi_{\vq}(i\omega_{n_1},i\omega_{n_2};i\omega_n)$,
\begin{widetext}
\begin{equation}
  \tilde\chi_{\vq}(i\omega_{n_1},i\omega_{n_2};i\omega_n)=
 \tilde\chi^0_{\vq}(i\omega_{n_1},i\omega_{n_2};i\omega_n) 
+\frac{1}{\beta}\sum_{n_3,n_4} 
\tilde\chi^0_{\vq}(i\omega_{n_1},i\omega_{n_3};i\omega_n)\Gamma^{(\rm
  pp)}(i\omega_{n_3},i\omega_{n_4};i\omega_n)
\tilde\chi_{\vq}(i\omega_{n_4},i\omega_{n_2};i\omega_n).
\label{chipp}
\end{equation}
\end{widetext}
$\Gamma^{(\rm  pp)}(i\omega_{n_1},i\omega_{n_4};i\omega_n)$ is the irreducible
vertex in the particle-particle channel which is local in
DMFT. The corresponding pair
propagator $\tilde\chi^0_{\vq}(i\omega_{n_1};i\omega_n) $ is 
\begin{equation}
  \tilde\chi^0_{\vq}(i\omega_{n_1},i\omega_{n_2};i\omega_n)=\sum_{\vk}G_{\vk}(i\omega_{n_1})G_{\vq-\vk}(i\omega_n-i\omega_{n_1})\delta_{n_1,n_2}, 
\end{equation}
For special values of $\vq$ and $i\omega_n$ we can evaluate the pair
propagator.\cite{GKKR96} We are interested in the limit $\vq\to 0$ and
$i\omega_n\to0$, and one finds 
\begin{equation}
  \tilde\chi^0_0(i\omega_{n_1},i\omega_{n_2};0)=\frac{G(i\omega_{n_1})-G(-i\omega_{n_1})}{\zeta(-i\omega_{n_1})-\zeta(i\omega_{n_1})}\delta_{n_1,n_2}, 
\label{tchi0gen}
\end{equation}
where $\zeta(i\omega_{n})=i\omega_{n}+\mu-\Sigma(i\omega_{n})$ 
and
\begin{equation}
  G(i\omega_n)=\integral{\epsilon}{}{}\frac{\rho_0}{\zeta(i\omega_n)-\epsilon}
\equiv {\rm
    HT}[\rho_0](\zeta(i\omega_n)).
\label{eq:Gnorm}
\end{equation}
One has for the semi-elliptic DOS 
\begin{equation}
  {\rm
    HT}[\rho_0](z)=\integral{\epsilon}{-D}{D}\frac{\rho_0(\epsilon)}{z-\epsilon}=\frac{1}{2
    t^2}\Big(z-{\rm sgn}(\Imag(z))\sqrt{z^2-4t^2}\Big),
\end{equation}
where the square root of a complex number $w$ is given by
$\sqrt{r}\e^{i\varphi/2}$, where $\varphi=[0,2\pi)$, such that the imaginary
part of $\sqrt{w}$ is positive. At half filling $G(i\omega_n)$ and
$\Sigma(i\omega_n)$ are purely imaginary functions. The phonon
Green's function $D(i\omega_m)$,
\begin{equation}
  D(i\omega_m)^{-1}=D^0(i\omega_m)^{-1}-\Sigma_{{\rm ph}}(i\omega_m),
\end{equation}
where $D^0(i\omega_m)=2\omega_0/[(i\omega_m)^2-\omega_0^2]$,
 and its self-energy $\Sigma_{{\rm  ph}}(i\omega_m)$ are real
 functions. In the non-interacting case  the Green's function reads, 
\begin{equation}
  G(i\omega_n)=\frac{i}{2t^2}\Big(\omega_n-{\rm
    sgn}(\omega_n)\sqrt{\omega_n^2+4t^2}\Big).
\label{Gfunc}
\end{equation}
At half filling $\mu=0$, $G(-i\omega_{n})=-G(i\omega_{n})$,
$\Sigma(-i\omega_{n})=-\Sigma(i\omega_{n})$. With
$\Sigma(i\omega_{n})=i\omega_n(1-Z(i\omega_n))$, where $Z(i\omega_n)$ is
symmetric and real, we find 
\begin{equation}
  \tilde\chi^0_0(i\omega_{n_1},i\omega_{n_2};0)=-\frac{G(i\omega_{n_1})}{i\omega_{n_1}Z(i\omega_{n_1})}\delta_{n_1,n_2}.
\label{symchi}
\end{equation}
This expression is positive. 

We can write Eq.~(\ref{chipp}) as a matrix equation (omitting the general arguments,
$\vq,i\omega_n$),
\begin{equation}
  \tilde\chi=\tilde\chi^0+\tilde\chi^0\Gamma\tilde\chi .
\end{equation}
The instability criterion is that $\chi$ diverges for some
$T,i\omega_n,\vq$.  This can be written as the
eigenvalue equation $[\tilde\chi^0 \Gamma] v=v$, or with $v=[\tilde\chi^0]^{1/2}
\tilde v$, 
\begin{equation}
[\tilde\chi^0]^{1/2}\Gamma [\tilde\chi^0]^{1/2}\tilde v=\tilde v ,
\label{eveq}
\end{equation}
where the matrix is symmetric if $\Gamma$ is symmetric.
The relevant symmetric matrix reads,
\begin{equation}
M_{n_1,n_2}= \frac{1}{\beta}\sqrt{\tilde\chi^0(i\omega_{n_1})}[\Gamma^{(\rm
  pp)}(i\omega_{n_1},i\omega_{n_2};0)]\sqrt{\tilde\chi^0(i\omega_{n_2})},
\label{symmatrixinst}
\end{equation}
and we have to find its largest eigenvalue. We have simplified the
notation for the arguments for $\tilde\chi^0$ omitted the $\vq=0$-label.
In a DMFT calculation for a given temperature $T$, we first determine
$\Sigma(i\omega_n)$ and the full particle-particle irreducible vertex $\Gamma^{(\rm
  pp)}(i\omega_{n_1},i\omega_{n_2};0)$. The pair propagator
$\tilde\chi^0(i\omega_{n_1})$ is obtained from Eq. (\ref{tchi0gen}). Then we 
can search for the largest eigenvalue of the symmetric matrix in Eq. (\ref{symmatrixinst})
for the instability criterion.

\subsection{Calculations in the superconducting phase}
We also perform calculations in the superconducting phase. We work in Nambu
space with matrices then. The local lattice Green's functions have the form 
\begin{equation}
G_{11}(i\omega_n)
= A_G {\rm HT}[\rho_0](\epsilon_+)+B_G {\rm HT}[\rho_0](\epsilon_-)
\end{equation}
and
\begin{equation}
G_{21}(i\omega_n)
= A_F {\rm HT}[\rho_0](\epsilon_+)+B_F {\rm HT}[\rho_0](\epsilon_-)
\label{eq:G21}
\end{equation}
with
$A_G=(\zeta_2(i\omega_n)+\epsilon_+(i\omega_n))/(\epsilon_+(i\omega_n)-\epsilon_-(i\omega_n))$,
$B_G=(\zeta_2(i\omega_n)+\epsilon_-(i\omega_n))/(\epsilon_-(i\omega_n)-\epsilon_+(i\omega_n))$
$A_F=\Sigma_{21}(i\omega_n)/(\epsilon_+(i\omega_n)-\epsilon_-(i\omega_n))$, and
$B_F=\Sigma_{21}(i\omega_n)/(\epsilon_-(i\omega_n)-\epsilon_+(i\omega_n))$, where
\begin{eqnarray*}
\epsilon_{\pm}&=&\frac{\zeta_1(i\omega_n)-\zeta_2(i\omega_n)}{2}\pm \\
&&\frac12\sqrt{(\zeta_1(i\omega_n)+\zeta_2(i\omega_n))^2-4\Sigma_{21}(i\omega_n)  \Sigma_{12}(i\omega_n)},
\end{eqnarray*}
with $\zeta_{1}(z)=z+\mu-\Sigma_{11}(z)$ and
$\zeta_{2}(z)=z-\mu-\Sigma_{22}(z)$.  We have
$G_{12}(i\omega_n)=G_{21}(i\omega_n)$ and
$G_{22}(i\omega_n)=-G_{11}(-i\omega_n)$ for the Nambu Green's functions.
We use $\Sigma_{12}(i\omega_n)=\Sigma_{21}(i\omega_n)$ and
$\Sigma_{22}(i\omega_n)=-\Sigma_{11}(-i\omega_n)$ for the self-energies. This can
be deduced from the properties of the corresponding Green's functions
including the assumption of time-reversal symmetry.
At half filling  $G_{11}(i\omega_n)$ and
$\Sigma_{11}(i\omega_n)$ are imaginary functions, whereas $G_{21}(i\omega_n)$ and
$\Sigma_{21}(i\omega_n)$ and $D(i\omega_m)$ and $\Sigma_{{\rm ph}}(i\omega_m)$ are real functions. 

In the NRG approach we calculate the self-energy matrix for the effective
impurity model from the matrix of higher Green's functions $\underline F(\omega)$
with
$F_{11}(\omega)=\gfbraket{\elann{d}{\uparrow}n_{\downarrow};\elcre{d}{\uparrow}}_{\omega}$,
$F_{12}(\omega)=\gfbraket{\elann{d}{\uparrow}n_{\downarrow};\elann{d}{\downarrow}}_{\omega}$, 
$F_{21}(\omega)=-\gfbraket{\elcre{d}{\downarrow}n_{\uparrow};\elcre{d}{\uparrow}}_{\omega}$ and
$F_{22}(\omega)=-\gfbraket{\elcre{d}{\downarrow}n_{\uparrow};\elann{d}{\downarrow}}_{\omega}$.
For the phonon part we use
$M_{11}(\omega)=\gfbraket{\elann{d}{\uparrow}(b+b^{\dagger});\elcre{d}{\uparrow}}_{\omega}$,
$M_{12}(\omega)=\gfbraket{\elann{d}{\uparrow}(b+b^{\dagger});\elann{d}{\downarrow}}_{\omega}$, 
$M_{21}(\omega)=-\gfbraket{\elcre{d}{\downarrow}(b+b^{\dagger});\elcre{d}{\uparrow}}_{\omega}$ and
$M_{22}(\omega)=-\gfbraket{\elcre{d}{\downarrow}(b+b^{\dagger});\elann{d}{\downarrow}}_{\omega}$.
In the NRG we calculate $M_{11}$ and $M_{21}$ directly on the real axis from
the Lehman spectral representation. The others follow from
$M_{12}(\omega)=-M_{21}(-\omega)^*$ and $M_{22}(\omega)=M_{11}(-\omega)^*$.
We can define the self-energy matrix by
\begin{equation}
  \underline \Sigma(\omega)=  U \underline F(\omega)\underline
  {G}(\omega)^{-1}+g \underline M(\omega)\underline
  {G}(\omega)^{-1}.
\label{SigFM}
\end{equation}
For self-consistency the local lattice Green's functions $\underline G(\omega)$ has
to be equal  to the impurity Green's function, $\underline\thGf(\omega)=G(\omega)$, where
\begin{equation}
\thGf^{-1}(\omega)=\omega\unitop_2+\mu\tau_3-  \underline K(\omega)- \underline \Sigma(\omega),
\end{equation}
with the matrix $\underline K(\omega)$ describing the effective medium.
We can take the form of the effective impurity model to correspond to an
Anderson-Holstein impurity model \cite{HM02} and calculations are carried out
as detailed, for instance in Ref.~\onlinecite{BHD09}. 
We solve the effective impurity problem with the numerical renormalization
group \cite{Wil75,BCP08} (NRG) adapted to the case with symmetry
breaking. The NRG has been shown to be very successful for calculating the local
dynamic response functions, and we use the approach\cite{PPA06,WD07}
based on complete basis set proposed by Anders and Schiller.\cite{AS05} For
the logarithmic discretization parameter we take the value  $\Lambda=1.8$ and
keep about 1000 states at each iteration. The initial bosonic Hilbert space is
restricted to a maximum of 50 states.  We will mainly consider two cases: (i)
constant density of states $\rho_0=1/W$, where $W$ is the bandwidth and (ii)
the semi-elliptic DOS $\rho_0(\epsilon)=\sqrt{4t^2-\epsilon^2}/(2\pi t^2)$
with $W=4t=2D$.

\section{Diagrammatic calculation for $T_c$}
\label{sec:diagrammaticsTc}

In this section we first show how the standard approach to conventional
superconductivity, ME theory, would be applied to the model under
consideration, and which diagrams are included. To determine $T_c$ we use the
instability criterion, Eq.~(\ref{eveq}), which is equivalent to the linearized
ME equations.
\subsection{Standard diagrammatics, ME theory}
\begin{figure}[!htpb]
\centering
\subfigure[]{\includegraphics[width=0.23\textwidth]{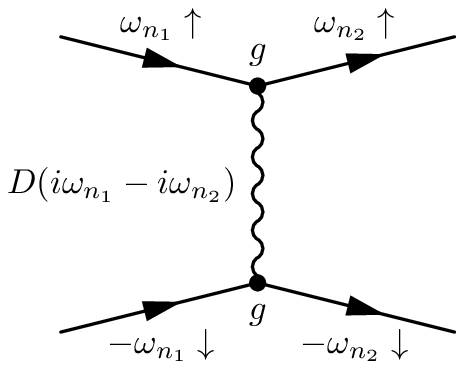}}
\hspace{0.5cm}
\subfigure[]{\includegraphics[width=0.2\textwidth]{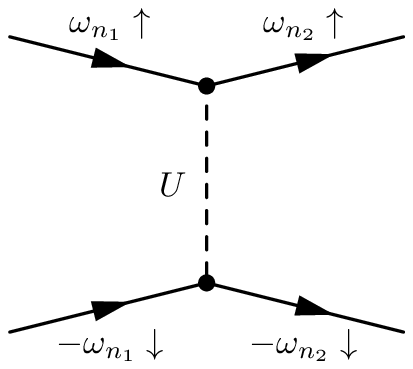}}
\caption{Contributions to the irreducible vertex (a) electron-phonon part, (b)
  first order in $U$.}       
\label{vertexfirstorder}
\end{figure}

In ME theory the irreducible vertex in
Eq. (\ref{symmatrixinst}) is given by the full phonon propagator  
\begin{equation}
  \Gamma^{(\rm
    pp)}(i\omega_{n_1},i\omega_{n_2};i\omega_n=0)=-g^2D(i\omega_{n_1}-i\omega_{n_2}).
\label{elphvert}
\end{equation}
Diagrammatically this is depicted in Fig.~\ref{vertexfirstorder} (a).

\begin{figure}[!htpb]
\centering
\subfigure[]{\includegraphics[width=0.22\textwidth]{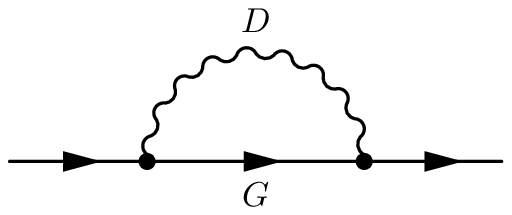}}
\hspace{0.5cm}
\subfigure[]{\includegraphics[width=0.22\textwidth]{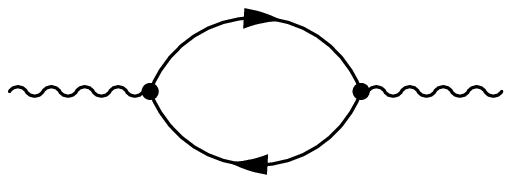}}
\caption{Diagrams for (a) electronic self-energy and (b) the phonon self-energy.}       
\label{MEdiagrams}
\end{figure}

\noindent
Due to Migdal's theorem \cite{Mig58} other vertex corrections including the
phonon propagator are
neglected. In general, the ME theory is a self-consistent calculation, and the
electronic self-energy reads [see Fig.~\ref{MEdiagrams} (a)], 
\begin{equation}
    \Sigma(i\omega_n)=-\frac{g^2}{\beta}\sum_{m}G(i\omega_m+i\omega_n)D(i\omega_m).
\label{eq:sigel}
\end{equation}

\noindent
The phonon self-energy reads [see Fig.~\ref{MEdiagrams} (b)],
\begin{equation}
   \Sigma_{{\rm
       ph}}(i\omega_m)=\frac{2g^2}{\beta}\sum_{n}G(i\omega_n)G(i\omega_m+i\omega_n).
\label{eq:sigphon}
\end{equation}
We define $\omega_0^r$ as the relevant renormalized phonon scale, which is
extracted from the peak position of the full phonon spectral function 
$\rho^D(\omega)$.  The electronic Green's function $G(i\omega_n)$ is
determined as in Eq.~(\ref{eq:Gnorm}). An appropriate
definition of the coupling constant $\lambda$ is 
\begin{equation} 
  \lambda=2\integral{\omega}{0}{\infty}\frac{\alpha^2F(\omega)}{\omega},
\end{equation}
where we the pairing function is defined via a Fermi surface average,
\begin{equation}
 \alpha^2F(\omega)=\frac1{\rho_0^2}\sum_{\vk,\vk'}
\alpha_{\vk,\vk'}^2F(\omega)
\delta(\epsilon_{\vk}-\mu)\delta(\epsilon_{\vk'}-\mu)
\end{equation}
with
\begin{equation}
 \alpha_{\vk,\vk'}^2F(\omega)=\rho_0|g_{\vk,\vk'}|^2\rho^D_{\vk-\vk'}(\omega).
\end{equation}
For the Holstein model using the free spectral function we have
$\alpha^2F(\omega)=\rho_0g^2[\delta(\omega-\omega_0)-\delta(\omega+\omega_0)]$,
such that we obtain $\lambda_0=\rho_02g^2/\omega_0$, purely in terms of bare
parameters. However, more relevant is the effective coupling $\lambda$ for the  
Holstein model is given by the full renormalized phonon propagator,\cite{BHG11}
\begin{equation}
  \lambda=2\rho_0g^2\integral{\omega}{0}{\infty}\frac{\rho^D(\omega)}{\omega}=-\rho_0g^2D(0). 
\label{eq:lambdadef1}
\end{equation}
The dimensionless quantity for the Coulomb interaction corresponding to
$\lambda_0$ is $\mu_c=\rho_0 U$.

In a self-consistent numerical calculation for a given temperature $T$, we
first determine $ \Sigma(i\omega_n)$ and $\Sigma_{{\rm
    ph}}(i\omega_m)$ by iterating Eqs.~(\ref{eq:sigel}) and (\ref{eq:sigphon}) and thus 
$G(i\omega_n)$ and $D(i\omega_m)$. Then we determine
$\tilde\chi^0(i\omega_{n_1})$ from Eq.~(\ref{tchi0gen}). $D(i\omega_m)$ is
used in Eq.~(\ref{elphvert}) to determine the irreducible vertex. Then we
calculate the largest eigenvalue of the symmetric matrix in Eq.~(\ref{symmatrixinst})
for the instability criterion. Instead of calculating the phonon Green's
function self-consistently we can also take it as an input.
Using the DMFT results for $D(i\omega_m)$  we found in previous work
\cite{BHG11} that $T_c$ obtained from this procedure agrees well with the
full DMFT result as long as the renormalized Migdal condition, $\lambda
\omega_0^r/W$ small, is satisfied. In the DMFT calculation  $\Gamma^{(\rm pp)}$
and $\Sigma(i\omega_n)$ contain all higher order corrections. 

In the usual theory the Coulomb repulsion is included directly up to first
order or as an additional empirical parameter $\mu^*$. The
vertex has then two contributions, one as before due to the electron-phonon 
interaction, and the second one from the Coulomb repulsion [see
Fig.~\ref{vertexfirstorder} (b)],
\begin{equation}
  \Gamma^{(\rm
    pp)}(i\omega_{n_1},i\omega_{n_2};i\omega_n=0)=-g^2D(i\omega_{n_1}-i\omega_{n_2})-U.
\label{eq:MEpairvert}
\end{equation}

The eigenvalue equation (\ref{eveq}) with the pairing vertex
(\ref{eq:MEpairvert}) can be solved analytically with approximations similar
to the ones by McMillan.\cite{McM68} This yields
\begin{equation}
  T_c=\frac{2 C\omega^r_0}{\pi}\exp\left[-\frac{Z}{\lambda-\mu_c^*[1+\frac12\lambda ]}\right],
\label{Tcapprox}
\end{equation}
with $C=\e^{\gamma}\approx 1.78$ with the Euler-Mascheroni constant $\gamma=0.57721$.
We have introduced $Z=1-\lim_{\omega\to 0} \Sigma(i\omega)/i\omega$ and
$\mu_c^*$ is given by Eq. (\ref{eq:mustar}) with $E_{\rm el}=D$ and $\omega_{\rm
  ph}=\omega_0^r$. 
This corresponds to the result by McMillan\cite{McM68} or  Allen and
Dynes \cite{AD75},
\begin{equation}
  T_c=\frac{\expval{\omega}{}}{1.2}\e^{-\frac{1.04(1+\lambda)}{\lambda-\mu_c^*(1+0.62\lambda
      )}},
\label{eq:ADeq}
\end{equation}
where $Z=1+\lambda$ is used.
The essential feature is that the Coulomb repulsion, which is the same
for all $i\omega_n$ is effectively reduced from $\mu_c$ to $\mu_c^*$ when
counteracting the electron-phonon attraction with strength $\lambda$. This
shows how retardation effects assist the electron-phonon induced
superconductivity by suppressing the detrimental effects due to the Coulomb
repulsion. 

In summary, the diagrams in Figs. \ref{vertexfirstorder} and \ref{MEdiagrams}
are the ones included in the standard theory of superconductivity. Usually,
the phonons are not calculated self-consistently but rather taken as an input,
for instance from DFT calculations or from experiment. Also $\mu^*_c$ is usually
not calculated but used as a fitting parameters. In the following we consider
higher order corrections to the standard approach.

\subsection{Higher order terms}
The perturbation expansion for two different interactions is rather
involved, since terms of each perturbation series as well as mixtures
can appear. In a skeleton expansion higher order contributions can be
grouped into the following  terms for self-energies and vertex functions:
\begin{enumerate}
\item
Contributions to the full electron-phonon vertex $\Gamma^{(\rm
  ep)}=\Gamma^{(\rm ep)}_g+\Gamma^{(\rm ep)}_U+\Gamma^{(\rm ep)}_{g,U} $,
where the bare vertex is $\Gamma^{(\rm ep)}_0=g$:
\begin{enumerate}
\item 
Higher order corrections $\Gamma^{(\rm ep)}_g$ purely due to $g$, not
including $\Gamma^{(\rm ep)}_0$.   
\item
Contributions $\Gamma^{(\rm ep)}_U$ which include the bare vertex $\Gamma^{(\rm ep)}_0$ and
higher order terms purely due to  $U$. 
\item
Higher order corrections $\Gamma^{(\rm ep)}_{g,U}$ due to mixed terms
of $g$ and $U$, not including $\Gamma^{(\rm ep)}_0$.  
\end{enumerate}
\item
Contributions to the phonon self-energy $\Sigma_{\rm ph}$:
\begin{enumerate}
\item
Higher order contributions to $\Sigma_{\rm ph}$, which can be written
in terms of $\Gamma^{(\rm ep)}_g$, i.e., purely due to $g$. 
\item
Contributions to $\Sigma_{\rm ph}$ due to mixed terms of $g$ and
$U$. These can be expressed in terms of $\Gamma^{(\rm ep)}_U$ or
$\Gamma^{(\rm   ep)}_{g,U}$. 
\end{enumerate}
\item
Contributions to the electron self-energy $\Sigma$:
\begin{enumerate}
\item
Contributions to $\Sigma$, which can be written
in terms of $\Gamma^{(\rm ep)}_g$, i.e. purely due to $g$. 
\item
Contributions to $\Sigma$ purely due to $U$. 
\item
Contributions to $\Sigma$ due to mixed terms, which can be
expressed in terms of $\Gamma^{(\rm ep)}_U$ or $\Gamma^{(\rm
  ep)}_{g,U}$.
\item
Contributions to $\Sigma$ due to mixed terms, which can not be written
in terms of $\Gamma^{(\rm ep)}$.
\end{enumerate}

\item
Contributions to the irreducible vertex $\Gamma^{(\rm pp)}$:
\begin{enumerate}
\item
Contributions to $\Gamma^{(\rm pp)}$ purely due to $U$. 
\item
Contributions to $\Gamma^{(\rm pp)}$ due to $g$ and $U$, which can be
written in terms of full propagators and the electron-phonon vertex $\Gamma^{(\rm ep)}$.
\item
Contributions to $\Gamma^{(\rm pp)}$ due to $g$ and $U$, which can not
be written in terms of $\Gamma^{(\rm ep)}$.
\end{enumerate}
\end{enumerate}
We assume in the following that the parameters are chosen such that
higher order contributions of the type $\Gamma^{(\rm ep)}_g$, i.e.,
1(a), 2(a) and 3(a), are small due to Migdal's theorem.
By taking the full phonon propagator from the DMFT
calculation as in previous work,\cite{BHG11} we avoid considering
in detail effects of 2(b), and rather assume that we can include the
correct phonon propagator. We focus on contributions 1(b), 3(b,c),  and 4(a)(b), in the
following, which are the main contributions in the low order
perturbation theory. Contributions to the type 1(c), 3(d) and 4(c) include
higher order diagrams and will not be considered  explicitely here.

\begin{figure}[!htpb]
\centering
\includegraphics[width=0.48\textwidth]{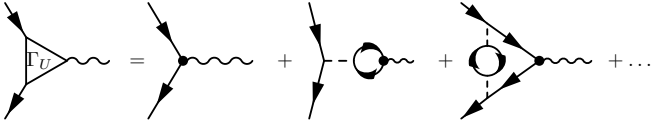}
\caption{Contributions to the electron-phonon vertex $\Gamma^{(\rm ep)}_U$.}       
\label{phverteqwU}
\end{figure}

We can generally write contributions to the pairing vertex of the form 4(b) as
\begin{widetext}
\begin{equation}
  \Gamma^{(\rm
    pp)}(i\omega_{n_1},i\omega_{n_2};i\omega_n=0)=
-\Gamma^{(\rm ep)}_U(i\omega_{n_1},i\omega_{n_2})D(i\omega_{n_1}-i\omega_{n_2})\Gamma^{(\rm ep)}_U(-i\omega_{n_1},-i\omega_{n_2}),
\label{eq:gammaU}
\end{equation}
\end{widetext}
where $\Gamma^{(\rm ep)}_U$ includes $g$ and all corrections due to
$U$, see Fig. \ref{phverteqwU}. 

\begin{figure}[!htpb]
\centering
\subfigure[]{\includegraphics[width=0.2\textwidth]{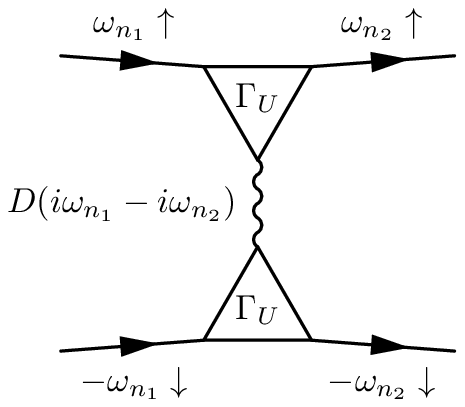}}
\hspace{0.5cm}
\subfigure[]{\includegraphics[width=0.22\textwidth]{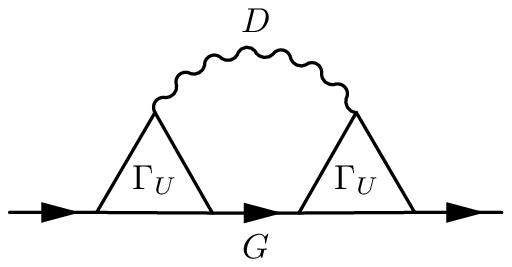}}
\caption{(a) Higher order terms involving vertex corrections of the
  electron-phonon vertex. (b) Electronic self-energy including vertex
  corrections for the electron-phonon vertex.}       
\label{correctelphaU}
\end{figure}

\noindent
For $\Gamma^{(\rm ep)}_U(i\omega_{n_1},i\omega_{n_2})$,
$i\omega_{n_1}$ is the ingoing electronic frequency, $i\omega_{n_2}$ the
outgoing one and the bosonic one is $i\omega_{n_1}-i\omega_{n_2}$. 
$D(i\omega_n)$ is the full propagator including corrections due to $U$
and $g$. Diagrammatically, these contributions to $\Gamma^{(\rm   pp)}$
are displayed in Fig. \ref{correctelphaU} (a).

\noindent
The higher order corrections included in Eq. (\ref{eq:gammaU}) can be 
seen as a redefinition of $\lambda$ defined in
(\ref{eq:lambdadef1}).  The vertex $\Gamma^{(\rm ep)}_U$ does not vary
much up to the small phonon scale, such that we can write approximately with
$g^r = \Gamma^{(\rm ep)}_U(0,0)$, 
\begin{equation}
  \lambda\simeq
  2\rho_0[g^r]^2\integral{\omega}{0}{\infty}\frac{\rho^D(\omega)}{\omega}=-\rho_0[g^r]^2D(0), 
\label{lambdafinU}
\end{equation}
where $\rho^D(\omega)$ can also include self-energy corrections due to
$U$. Such a redefined $\lambda$ enters the approximate results for $T_c$ such
as Eq. (\ref{Tcapprox}).
We can also take into account the effect that $\Gamma^{(\rm ep)}_U$ is a function of
frequency and average over a range given by the phonon Green's function such that
\begin{equation}
  [g^r]^2=\frac{1}{N_D} \sum_{n_2}\Gamma^{(\rm ep)}_U(0,i\omega_{n_2})^2D(i\omega_{n_2}),
\end{equation}
where $N_D=\sum_{n_2}D(i\omega_{n_2})$.

We now consider the diagrams contributing to the electron-phonon vertex $\Gamma^{(\rm ep)}_U$. Some
of them are shown in Fig. \ref{phverteqwU}.
To order $Ug$ we have
\begin{equation}
  \Gamma^{(\rm ep, 1)}_U(i\omega_{n_1},i\omega_{n_2})=g[1+U\Pi(i\omega_{n_1}-i\omega_{n_2})],
\label{elphvertcorr}
\end{equation}
where the particle-hole bubble is given by
\begin{equation}
\Pi(i\omega_n)= \frac{1}{\beta} \sum_{m} G(i\omega_n+i\omega_{m})G(i\omega_{m}).
\label{pidef}
\end{equation}
$\Pi(0)$ can be evaluated analytically at $T=0$ and one finds for the constant
density of states $\Pi(0)=-2\log(2)\rho_0$ and for the semi-elliptic DOS
$\Pi(0)=-\frac43\rho_0$. It turns out that a good empirical form for
$\Pi(i\omega)$ is given by  
\begin{equation}
  \Pi(i\omega)=\frac{-a\rho_0}{1+b_1|\omega|+b_2\omega^2}.
\label{pisimpform2}
\end{equation}
If $G(i\omega_n)$ is the non-interacting Green's function in Eq. (\ref{pidef})
then $a=4/3$ for the semi-elliptic DOS. In this case we find that $b_1=1.3/D$
and $b_2=1.63/D^2$ give a good fit. 

We can sum up a whole series of diagrams of the type discussed in
Eq. (\ref{elphvertcorr}) which corresponds to screening on the RPA level,
\begin{equation}
    \Gamma^{({\rm ep, RPA})}_U(i\omega_{n_1},i\omega_{n_2})=\frac{g}{1-U\Pi(i\omega_{n_1}-i\omega_{n_2})}.
\label{eq:gamscrrpa}
\end{equation}
Note the absence of a factor 2 in the denominator for the Hubbard interaction. 
Since $\Pi<0$ these diagrams lead to an effective reduction of the
electron-phonon coupling. If one assumes that $\Pi$ does not change much on
the phonon scale enforced via $D(i\omega_m)$, then one could approximate
$g^r=g/[1-U\Pi(0)]$, and for the Bethe lattice $g^r=g/[1+4\mu_c/3]$.

\begin{figure}[!htpb]
\centering
\subfigure[]{\includegraphics[width=0.12\textwidth]{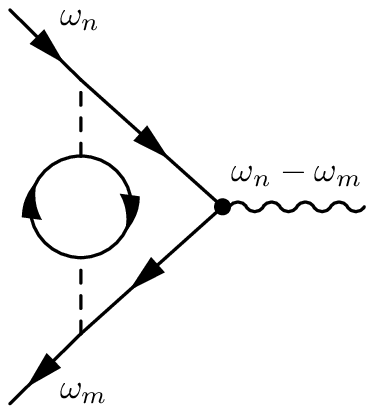}}
\hspace{0.5cm}
\subfigure[]{\includegraphics[width=0.12\textwidth]{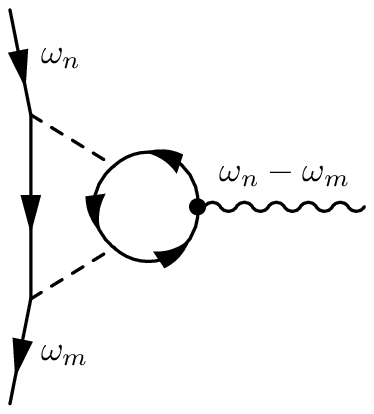}}
\hspace{0.5cm}
\subfigure[]{\includegraphics[width=0.12\textwidth]{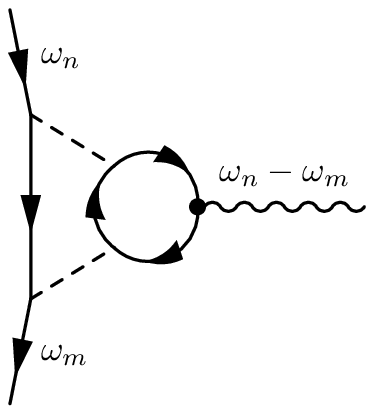}}
\caption{Vertex corrections to the electron-phonon vertex to order $gU^2$.}       
\label{correctelphgU2}
\end{figure}

There are three diagrams to order $U^2$ in addition to the screening term to
correct the vertex (see the third diagram in the 
Fig. \ref{phverteqwU} and Fig. \ref{correctelphgU2}). 

\noindent
The first term, Fig. \ref{correctelphgU2} (a), reads,
\begin{eqnarray}
  \Gamma^{(\rm ep,2,1)}_U(i\omega_n,i\omega_m)&=&-\frac{gU^2}{\beta}
  \sum_{k}\Pi(i\omega_k)G(i\omega_k+i\omega_n)\times \nonumber \\
&& \times G(i\omega_k+i\omega_m).
\label{eq:gammaUso}
\end{eqnarray}
The second one,  Fig. \ref{correctelphgU2} (b), is of the form
\begin{eqnarray}
  \Gamma^{(\rm ep,2,2)}_U(i\omega_n,i\omega_m)&=&-\frac{gU^2}{\beta}
  \sum_{k}\Pi_{pp}(i\omega_k)G(i\omega_k-i\omega_n)\times \nonumber \\
&& \times G(i\omega_k-i\omega_m), 
\end{eqnarray}
where we have introduced the particle-particle bubble,
\begin{equation}
\Pi_{pp}(i\omega_n)= \frac{1}{\beta} \sum_{m} G(i\omega_n-i\omega_{m})G(i\omega_{m}).
\end{equation}
At half filling we have $G(-i\omega_{m})=-G(i\omega_{m})$, which implies
$\Pi_{pp}(i\omega_n)=-\Pi(i\omega_n)$. With this one obtains
\begin{equation}
  \Gamma^{(\rm ep,2,2)}_U(i\omega_n,i\omega_m)=-\Gamma^{(\rm ep,2,1)}_U(i\omega_n,i\omega_m), 
\end{equation}
and the first diagram is canceled. The third diagram, Fig. \ref{correctelphgU2} (c), is like the first one
$\Gamma^{(\rm ep,2,3)}_U(i\omega_n,i\omega_m)=\Gamma^{(\rm ep,2,1)}_U(i\omega_n,i\omega_m)$. So
altogether we have just one contribution of the form (\ref{eq:gammaUso}) at
half filling. Diagrams of these type together with the RPA screening series
were discussed 
by Huang et al. \cite{HHAS03} for the two dimensional Hubbard-Holstein
model. The full electron-phonon vertex was calculated with QMC. It was found
that for values of $U$ up to half the bandwidth the diagrammatics and the
non-perturbative result are in good agreement. However, for
large values of $U$, the perturbation theory breaks down.

\begin{figure}[!htpb]
\vspace*{0.5cm}
\centering
\includegraphics[width=0.4\textwidth]{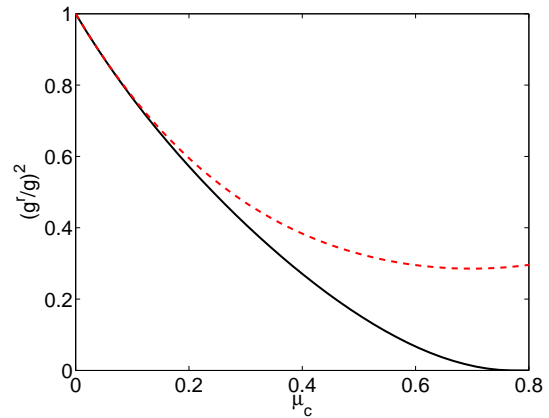}
\caption{(Color online) $(g^r/g)^2$ as a function of
  $\mu_c$, full line according to Eq. (\ref{eq:effvertcorr}) and
  dashed line according to Eq. (\ref{eq:effvertcorr2}) .}       
\label{fig:gr0g_mucdep}
\end{figure}

As can be seen from this analysis of the vertex correction, there is a
considerable effect to suppress the coupling $g$
when $U$ is finite. Including diagrams in Eq. (\ref{eq:gamscrrpa}) and
(\ref{eq:gammaUso}) we can write the approximate form,
\begin{equation}
  \frac{g^r}{g}=\frac{1}{1+a\mu_c}-b\mu_c^2,
\label{eq:effvertcorr}
\end{equation}
where $a=-\Pi(0)/\rho_0$ and 
\begin{equation}
b=\frac{1}{\rho_0^2\beta} \sum_{k}\Pi(i\omega_k)G(i\omega_k)^2.
\end{equation}
In the simplest case for free Green's functions and a semi-elliptic
DOS we have $a=4/3$ and $b=0.8237$ numerically. In a calculation
strictly up to second order in $U$ we have instead of
Eq. (\ref{eq:effvertcorr}), 
\begin{equation}
  \frac{g^r}{g}=1-a\mu_c+(a^2-b)\mu_c^2,
\label{eq:effvertcorr2}
\end{equation}
We plot results for $(g^r/g)^2$ in Fig. \ref{fig:gr0g_mucdep}.  

\noindent
We see that there occurs a substantial suppression, such that
already for $\mu_c=0.25$ the quantity $(g^r/g)^2$ is reduced to about
a half of its value for $\mu_c=0$. Since $a^2>b$, the second term in
Eq. (\ref{eq:effvertcorr2}) which dominates for larger $\mu_c$, leads
to an upturn of the result. 
The result in Eq. (\ref{eq:effvertcorr}) overestimate the suppression effect
for large values of $\mu_c$.\cite{HHAS03} 
For the interacting system, the coefficients
$a$ and $b$ tend to be smaller than the values in the non-interacting
limit. 

The vertex corrections $\Gamma^{(\rm ep)}_U$ also enter the electronic
self-energy as one class of diagrams contributing to 3(c). This can be
written as 
\begin{widetext}
\begin{equation}
    \Sigma(i\omega_n)=-\frac{1}{\beta}\sum_{m}\Gamma^{(\rm ep)}_U(i\omega_n,i\omega_m+i\omega_n)G(i\omega_m+i\omega_n)D(i\omega_m)
    \Gamma^{(\rm ep)}_U(i\omega_m+i\omega_n,i\omega_n).
\label{eq:elselfmixed}
\end{equation}
\end{widetext}
This is shown diagrammatically in Fig. \ref{correctelphaU} (b).
As discussed above, there are also other types of mixed diagrams 3(d),
which cannot be written in the form of
Eq. (\ref{eq:elselfmixed}). 

\subsection{Higher order corrections from purely $U$}
\begin{figure}[!htpb]
\centering
\subfigure[]{\includegraphics[width=0.22\textwidth]{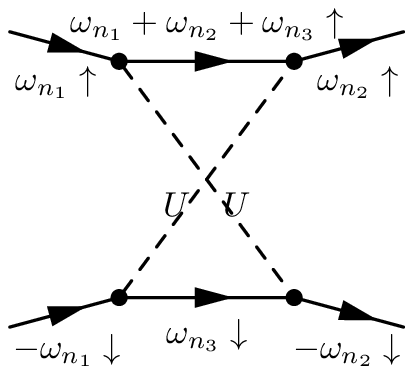}}
\hspace{0.5cm}
\subfigure[]{\includegraphics[width=0.22\textwidth]{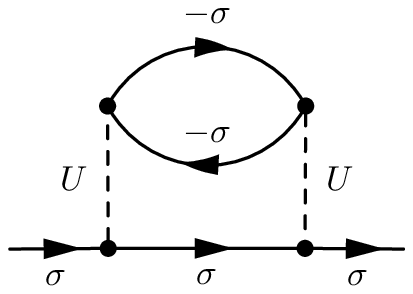}}
\caption{(a) Higher order diagram vertex contribution from the Coulomb
  repulsion. (b) Second order in $U$ diagram for electronic self-energy.}       
\label{elselfenU}
\end{figure}

We now deal with the higher order corrections purely from $U$, i.e.,
of type 4(a).  
We will restrict our attention only to the terms second order in $U$. 
The corresponding contribution (crossed diagram) to the pairing vertex reads
\begin{equation}\label{eq:o00}
  \Gamma^{(\rm pp)}(i\omega_{n_1},i\omega_{n_2};i\omega_n=0)= U^2 \Pi(i\omega_{n_1}+i\omega_{n_2}).
\end{equation}
The diagram is depicted in Fig. \ref{elselfenU} (a). 

\noindent
A naive way of taking this into consideration would be to assume that
$\Pi(i\omega_{n_1}+i\omega_{n_2})$ varies little for the scales under
considerations, such that for all frequencies we can assume
$-\rho_0U^2\Pi(0)=a\mu_c^2$. Then this term can be treated in the same way as
the first order term in $U$, where we can simply write
$\mu_c\to\bar\mu_c=\mu_c+a\mu_c^2$. This quantity is then subject to the
full retardation effects and becomes $\bar\mu_c^*$,
\begin{equation}
  \bar\mu_c^*=\frac{\bar\mu_c}{1+\bar\mu_c\log(\frac{D}{\omega_{\rm ph}})},
\label{eq:mustarbar}
\end{equation}
However, as analyzed in detail in Sec.~\ref{sec:resmustar} retardation effects are less efficient
for the second order term when the decay of $\Pi(i\omega)$ is taken into account properly. 

For the self-energy contributions of type 3(b), we will only consider
the standard second order diagram depicted in Fig. \ref{elselfenU} (b),
\begin{equation}
  \Sigma(i\omega_n)=-\frac{U^2}{\beta^2}\sum_{n_1,n_2}
 G(i\omega_n-i\omega_{n_1})G(i\omega_{n_1}+i\omega_{n_2})G(i\omega_{n_2}).
\label{eq:sodiagU}
\end{equation}

\section{Diagrammatic calculation for the superconducting state}

In this section we consider calculations in the superconducting state. We
first present the application of the standard theory to the
HH model and then discuss higher order corrections.
These calculations allow us, for instance, to study the gap at $T=0$. The
relevant quantities are matrices in Nambu space. We have to calculate the
diagonal and off-diagonal components of the self-energy. First 
we present the diagrams usually used in the standard ME approach. 

\subsection{Standard diagrammatics, ME theory}
\begin{figure}[!t]
\centering
\subfigure[]{\includegraphics[width=0.22\textwidth]{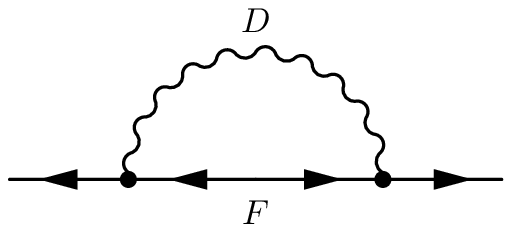}}
\hspace{0.5cm}
\subfigure[]{\includegraphics[width=0.22\textwidth]{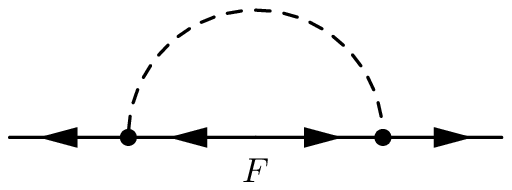}}
\caption{Diagrams for the off-diagonal self-energy (a) from electron-phonon
  interaction and (b) for the Coulomb repulsion with the notation $F=G_{21}$.}       
\label{offdiagdiag}
\end{figure}

To lowest order the diagonal self-energy is given by
\begin{equation}
    \Sigma_{11}(i\omega_n)=-\frac{g^2}{\beta}\sum_{m}G_{11}(i\omega_m+i\omega_n)D(i\omega_m),
\label{eq:MEsig11}
\end{equation}
and the off-diagonal self-energy reads,
\begin{equation}
    \Sigma_{21}(i\omega_n)=\frac{g^2}{\beta}\sum_{m}G_{21}(i\omega_m+i\omega_n)D(i\omega_m).
\label{eq:MEsig21}
\end{equation}
This is depicted in Fig. \ref{offdiagdiag} (a) with the off-diagonal
self-energy marked as $F$.

\noindent
In the limit of low temperature, $T \to 0$, we use 
\begin{equation}
  \frac1{\beta}\sum_nf(i\omega_n)\to
  \frac{1}{2\pi}\integral{\omega}{-\infty}{\infty}f(i\omega). 
\end{equation}
For the off-diagonal self-energy [see Fig. \ref{offdiagdiag} (b)] we
have the Coulomb contribution ($\eta\to0$)
\begin{equation}
    \Sigma_{21}(i\omega_n)=\frac{U}{\beta}\sum_{m}G_{21}(i\omega_m)\e^{i\omega_m\eta}.
\label{eq:sig21u1}
\end{equation}
Since $G_{21}(i\omega_m)$ decays sufficiently rapidly the factor
$\e^{i\omega_m\eta}$ is usually not needed.  

These are the equations, which are taken into account in the standard
theory. Often, the Coulomb term is then projected onto the phonon scale
(Coulomb pseudopotential), where only the reduced value $\mu_c^*$ enters. 
Making a few simplifying assumption (see Sec. \ref{sec:analyticgapT0}) we can derive for the spectral gap at
$T=0$, $\Delta_{\rm sp}\simeq \Sigma_{21}(0)/Z$, 
\begin{equation}
  \Delta_{\rm sp} = 2 \omega^r_0\e^{-\frac{Z}{\lambda-\mu_c^*(1+\frac{\lambda}{2})}}.
\end{equation}
This result is very similar to the one for $T_c$. Equations
(\ref{eq:MEsig11}), (\ref{eq:MEsig21}), and(\ref{eq:sig21u1}) correspond to
the standard approach for the theory of conventional superconductivity.\cite{Car90,MC08}

\subsection{Higher order corrections}
In a similar way as for the analysis of the pairing vertex in
Sec. \ref{sec:diagrammaticsTc}, we can classify terms for the
self-energy into different contributions. We will follow the same
logic as in the  calculation for $T_c$ and focus on the same type of
diagrams as before. Here we consider a subclass of type 3(c), which
includes Coulomb vertex corrections of the electron-phonon vertex of
the type $\Gamma^{(\rm ep)}_U$. It reads,
\begin{widetext}
\begin{equation}
    \Sigma_{11}(i\omega_n)=-\frac{1}{\beta}\sum_{m}\Gamma^{(\rm ep)}_U(i\omega_n,i\omega_m+i\omega_n)G_{11}(i\omega_m+i\omega_n)D(i\omega_m)
    \Gamma^{(\rm ep)}_U(i\omega_m+i\omega_n,i\omega_n).
\label{eq:sig11withvertex}
\end{equation}
Assuming a slow variation of $\Gamma_U$ on the phonon, scale a good
approximation is of the form, 
\begin{equation}
    \Sigma_{11}^{}(i\omega_n)\simeq-\frac{[g^r]^2}{\beta}\sum_{m_1}
 D(i\omega_{m_1})G_{11}(i\omega_n+i\omega_{m_1}).
\end{equation}
Similarly we have for the off-diagonal self-energy,
\begin{equation}
    \Sigma_{21}(i\omega_n)=\frac{1}{\beta}\sum_{m}\Gamma^{(\rm ep)}_U(i\omega_n,i\omega_m+i\omega_n)G_{21}(i\omega_m+i\omega_n)D(i\omega_m)
    \Gamma^{(\rm ep)}_U(i\omega_m+i\omega_n,i\omega_n).
\label{eq:sig21withvertex}
\end{equation}
\end{widetext}
For $\Gamma^{(\rm ep)}_U(i\omega_n,i\omega_m)$ we consider the same diagrams as in
Sec. \ref{sec:diagrammaticsTc} B. We assume that $D(i\omega_m)$ is the full
phonon propagator and hence it contains corrections due to $U$ as well. 

\begin{widetext}
\subsection{Higher order corrections from purely $U$}
Of the contributions of the type 3(b) we discuss all terms to second
order in $U$. In the  Nambu perturbation theory ones has
(cf. Ref. \onlinecite{MF92}),
\begin{equation}
    \Sigma_{11,U}^{(2,1)}(i\omega_n)=-\Big(\frac{U}{\beta}\Big)^2\sum_{m_1,m_2}
 G_{11}(i\omega_n+i\omega_{m_1})
 G_{22}(i\omega_{m_2})G_{22}(i\omega_{m_1}+i\omega_{m_2}).
\label{eq:sig11u2a}
\end{equation}
and
\begin{equation}
    \Sigma_{11,U}^{(2,2)}(i\omega_n)=\Big(\frac{U}{\beta}\Big)^2\sum_{m_1,m_2}
 G_{21}(i\omega_n+i\omega_{m_1})
 G_{12}(i\omega_{m_2})G_{22}(i\omega_{m_1}+i\omega_{m_2}).
\label{eq:sig11u2b}
\end{equation}
For the off-diagonal part we have 
\begin{equation}
    \Sigma_{21,U}^{(2,1)}(i\omega_n)=-\Big(\frac{U}{\beta}\Big)^2\sum_{m_1,m_2}
 G_{21}(i\omega_n+i\omega_{m_1})G_{12}(i\omega_{m_2})G_{12}(i\omega_{m_1}+i\omega_{m_2}), 
\label{eq:sig21u2a}
\end{equation}
\begin{equation}
    \Sigma_{21,U}^{(2,2)}(i\omega_n)=\Big(\frac{U}{\beta}\Big)^2\sum_{m_1,m_2}
 G_{11}(i\omega_n+i\omega_{m_1})
 G_{22}(i\omega_{m_2})G_{12}(i\omega_{m_1}+i\omega_{m_2}). 
\label{eq:sig21u2b}
\end{equation}
\end{widetext}
The first diagram $\Sigma_{11,U}^{(2,1)}(i\omega_n)$ is the well known
$U^2$-diagram which gives the first dynamic correction, also in the normal
phase as 
in Eq. (\ref{eq:sodiagU}). In comparison
$\Sigma_{11,U}^{(2,2)}(i\omega_n)$ gives a smaller contribution as it is
proportional to two off-diagonal Green's
function. $\Sigma_{21,U}^{(2,1)}(i\omega_n)$ is also comparably small, but
$\Sigma_{21,U}^{(2,2)}(i\omega_n)$ gives a sizeable 
reduction of the superconducting state. We can see this by writing it as
\begin{equation}
    \Sigma_{21,U}^{(2,2)}(i\omega_n)=-U^2\frac{1}{\beta}\sum_{m_1}\Pi(i\omega_n+i\omega_{m_1})G_{21}(i\omega_{m_1}).
\label{eq:sig21u2bmod}
\end{equation}
where $\Pi(i\omega_n)$ is given in Eq. (\ref{pidef}) with $G(i\omega_{m})=G_{11}(i\omega_{m})$.

For small $i\omega_n$ a crude approximation is to write
\begin{equation}
    \Sigma_{21,U}^{(2,2)}(0)\simeq -U^2c_1\Pi(0)\frac{1}{\beta}\sum_{m_1}G_{21}(i\omega_{m_1}),
\end{equation}
where it is assumed that $G_{21}(i\omega_{m_1})$ is only finite in small
interval such that we can take $\Pi(i\omega_n)$ constant.
Hence we find a direct correction to the term $\Sigma_{21,U}^{(1)}$. With
the result for $\Pi(0)$,  one has then approximately 
\begin{equation}
\Sigma_{21,U}^{(1)}+ \Sigma_{21,U}^{(2,2)}(0)\simeq U(1+
a\mu_c)\frac{1}{\beta}\sum_{m_1}G_{21}(i\omega_{m_1}),
\end{equation}
This is the same effect that was discussed for $T_c$ and the crossed
diagram [see Fig. \ref{elselfenU} (a)] where naively the effective $\mu_c$ becomes
$\bar\mu_c=\mu_c(1+a\mu_c)$. In a more accurate treatment, we have to take the frequency
dependence into account, and we will see that this leads to modifications.

\section{Analytic and numerical results for $\mu_c^*$}
\label{sec:resmustar}
In the last two sections we have analyzed the diagrammatic expansion
for the model in Eq. (\ref{hubholham}) and discussed certain types of
diagrams. In this section we want to  specifically study the pseudopotential
effect without including all other corrections.  
We are interested whether the first and second order calculations give
qualitatively different results for $\mu_c^*$. We present a
combination of analytical and numerical arguments.
In the literature, there exist a number of ways to calculate $\mu_c^*$
for a given microscopic model. 
One early approach by Bogoliubov et al. is based on an integral equation for the Coulomb
part in the pairing equation.\cite{BTS62,schrieffer}
Morel and Anderson \cite{MA62} gave an approximate solution for the
Migdal-Eliashberg equations including a screened Coulomb repulsion
in the $T=0$ formalism. The pseudopotential $\mu_c^*$ also appears
naturally when superconducting pairing instabilities are studied in a
renormalization group framework.\cite{Sha94,TTCC07}
In the following we first calculate $\mu_c^*$ directly by projecting
the pairing matrix to low energy. Analytic and numerical results are
compared. Then in a second approach we calculate $\Delta_{\rm sp}$
from an approximate solution of the self-consistency equation and thus
extract $\mu_c^*$. Finally we also compute numerical results for $T_c$
obtained from the instability condition and analyze these results in
terms of $\mu_c^*$.

\subsection{Projection scheme }

The starting point for the projection approach is the pairing matrix in 
the form 
\begin{equation}
\label{eq:o00000}
A_{nm}=\delta_{nm}-M_{nm},
\end{equation}
where $M_{nm}$ is given in Eq. (\ref{symmatrixinst}).
This matrix becomes singular at $T_c$.
We introduce the ``low-energy part'' 
\begin{equation}
\label{eq:o0}
  A^{\rm low}_{nm}=A_{nm}-\sum_{|\omega_n'|,|\omega_m'|>\omega_{\rm ph}}A_{nn'}[\bar A^{-1}]_{n'm'}A_{m'm},
\end{equation}
$n,m$ such that $|\omega_n|,|\omega_m|<\omega_{\rm ph}$, and $\bar A$ is the
matrix that is left after the blocks given by $|\omega_n|<\omega_{\rm ph}$ or
$|\omega_m|<\omega_{\rm ph}$ were removed. If $\bar A^{-1}$ is not singular,
Eq.~(\ref{eq:o00000}) and Eq.~(\ref{eq:o0}) 
become singular for the same parameters. This way we can reduce the matrix size so that it 
only includes frequencies for which the electron-phonon interaction is important.
The ``folding in'' of larger frequencies then describes how retardation effects 
reduce the effects of the Coulomb repulsion on low frequency properties.

We first consider the lowest order term of $\Gamma^{(\rm pp)}$ in $U$
namely $\Gamma^{(\rm pp)}=-U$. 
We want to focus on the dependence on the half-band width $D$ and assume 
that the density of states scales as
\begin{equation}\label{eq:o4}
\rho(\varepsilon)={1\over 2D}\bar\rho\Big({\varepsilon\over 2D}\Big),
\end{equation}
where $\bar\rho$ is independent of $D$. For simplicity we assume in the following that
$\bar\rho(x)=1$ is a constant for $|x|\le 1/2$ and zero
otherwise, such that $\rho_0=1/(2D)$.
It is then a rather good approximation to write 
\begin{equation}\label{eq:o0000}
\tilde \chi^0(i\omega_n)=
\begin{cases}
\rho_0\pi/|\omega_n|, & \text{if $|\omega_n|<D$};
\\  0,& \text{otherwise.}\end{cases}
\end{equation}  
The matrix $A$ then takes the form
\begin{equation}
\label{eq:o1} 
A_{nm}=\delta_{nm}+\frac{\pi }{\beta \sqrt{|\omega_n \omega_m|}}\mu_c,
\end{equation}
which is separable and can be inverted exactly. We obtain
\begin{equation}
\label{eq:o2}
[\bar A^{-1}]_{nm}=\delta_{nm}-\frac{\pi \mu_c}{\beta \sqrt{|\omega_n \omega_m|}
[1+\pi \mu_c/\beta\sum_k^{'} 1/|\omega_k|]},
\end{equation}
where $\sum_k^{'}$ involves a summation over $|\omega_k|>\omega_{\rm ph}$.
Replacing the summations in Eqs.~(\ref{eq:o0}, \ref{eq:o2}) by integrals, 
we find   
\begin{eqnarray}
\label{eq:o3}
&&A_{nm}^{\rm low}=\delta_{nm}+\frac{\pi}{\beta \sqrt{|\omega_n \omega_m|}}\mu_c \nonumber\\
&&-\frac{\pi }{\beta \sqrt{|\omega_n \omega_m|}} \frac{\mu_c^2{\rm log} 
(D/\omega_{\rm ph})}{1+\mu_c{\rm log}(D/\omega_{\rm ph})} \\            
&&=\delta_{nm}+\frac{\pi }{\beta \sqrt{|\omega_n \omega_m|}}
\frac{ \mu_c}{1+\mu_c{\rm log}(D/\omega_{\rm ph})}. \nonumber
\end{eqnarray}
Comparison of Eq. (\ref{eq:o1}) and (\ref{eq:o3}) leads to the Coulomb
pseudopotential, $\mu_c\to\mu_c^*$, as given in in Eq.~(\ref{eq:mustar}), and
it is the result of Morel and Anderson.\cite{MA62}

We next consider the second order term of $\Gamma^{(\rm pp)}$ in $U$ [Eq.~(\ref{eq:o00})]. 
Because of the form of $\Pi(i\omega_n+i\omega_m)$ it is then not possible to invert
$\bar A$ in Eq.~(\ref{eq:o0}) analytically. 
For the density of states [Eq.~(\ref{eq:o4})] we can rewrite $\Pi(i\omega_n+i\omega_m)$ as
\begin{equation}
\label{re:o5}
\Pi(i\omega_n+i\omega_m)=-{1\over 2D}f\Big(\frac{\omega_n+\omega_m}{2D}\Big),
\end{equation}
where $f$ is independent of $D$. As in Eq. (\ref{pisimpform2}) it is
quite accurate to approximate $f$ as
\begin{equation}
\label{eq:o6}
f(x)=\frac{a}{1+b|x|+cx^2},
\end{equation}
where $a=1.38$, $b=2$ and $c=5$ are suitable values for the constant DOS.

We now want to calculate $A^{\rm low}$ in Eq.~(\ref{eq:o0}) to third 
order in $U$. We use the result for $\bar A^{-1}$ in Eq.~(\ref{eq:o2}), which    
neglects the second order term and is therefore only correct to order $U$.
However, since the off-diagonal terms of $A$ in Eq.~(\ref{eq:o0}) are of the 
order $U$, the final result is of the order $U^3$. We obtain
\begin{eqnarray}
\label{eq:o7}
&&\sum_{nm}A_{1n}[\bar A^{-1}]_{nm}A_{m1}    \nonumber  \\ 
&&=\tilde \chi^0(i\omega_1)\Big({U\over \beta}\Big)^2\sum_{|\omega_n|\ge \omega_{\rm ph}}
\tilde\chi^0(i\omega_n)\Big[1+\mu_cf\Big({\omega_n \over 2D}\Big)\Big]^2 \nonumber \\
&&-\Big({U\over \beta}\Big)^3 {\tilde \chi^0(i\omega_1) \over 1+\mu_c{\rm log }(D/\omega_{\rm ph})}
  \\
&& \times \Big\lbrace \sum_{|\omega_n|\ge \omega_{\rm ph}}\tilde\chi^0(i\omega_n)
\Big[1+\mu_cf\Big({\omega_n \over 2D}\Big) \Big]\Big\rbrace ^2,
\nonumber
\end{eqnarray}
where we focus on the diagonal result for the lowest frequency $\omega_1$.
Using the definition of $\tilde\chi^0(i\omega_n)$ in Eq.~(\ref{tchi0gen}), not the approximation
in Eq.~(\ref{eq:o0000}), and applying the limit $\omega_{\rm ph}\ll D$, we calculate the sums 
\begin{equation}
\label{eq:o8}
\sum_{|\omega_n|\ge \omega_{\rm ph}}\tilde\chi^0(i\omega_n)f\Big({\omega_n \over 2D}\Big)^k=
{a^k\beta \over 2D}{\rm log}\Big({DA_k\over \omega_{\rm ph}}\Big),
\end{equation}
where $A_0\approx 1.00$, $A_1\approx 0.32$ and $A_2\approx 0.20$. For $k=1$ and 2
the denominator of $f$ reduces the integral for large $|\omega_n|$, which effectively 
reduces the band width by a factor $A_k$. This reduction is naturally larger
for $k=2$ than for $k=1$.

We now make an ansatz for $\mu_c^*$
\begin{equation}
  \mu_c^*=\frac{\mu_c+a\mu_c^2 }{1+\mu_c \log\Big(\frac{D}{\omega_{\rm
        ph}}\Big)+a\mu_c^2 \log\Big(\frac{D}{\omega_{\rm ph}}A_{22}\Big)}, 
\label{eq:mustarproj} 
\end{equation}
along the lines of the Morel-Anderson form, but including a second order term
and a corresponding logarithm. Based on Eq.~(\ref{eq:o8}) we expect the effective band
width to be smaller for the second order term and we therefore allow  for a different multiplying
factor $A_{22}$ in the logarithm. Since the result in Eq.~(\ref{eq:o7}) is correct to order 
$U^3$ the factor $A_{22}$ in Eq.~(\ref{eq:mustarproj}) can be identified, $A_{22}=A_1^2/A_0
\approx 0.10$. The ansatz in Eq.~(\ref{eq:mustarproj}) is then also correct to
order $U^3$.

\begin{figure}[!htpb]
\centering
\includegraphics[width=0.3\textwidth,angle=270]{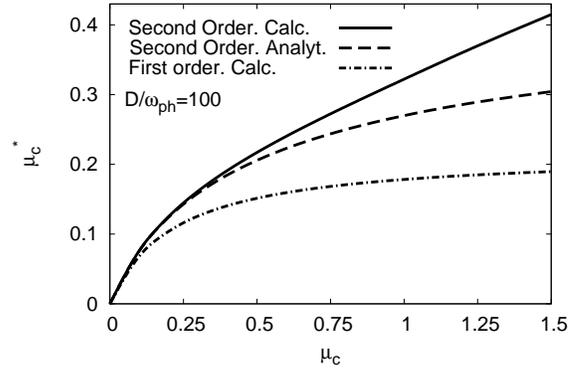}
\caption{The pseudopotential $\mu^*_c$ as a function of $\mu_c$ for $D/\omega_{\rm ph}=100$ and $\beta \omega_{\rm ph}=240$.
The figure shows the calculated results using both the first order and first plus second
order result for $\Gamma^{(\rm pp)}$ as well as the approximation in 
Eq.~(\ref{eq:mustarproj}).  }
\label{fig:mustarproj}
\end{figure}

Fig. \ref{fig:mustarproj} shows the results obtained by performing the calculations
in Eq.~(\ref{eq:o0}) numerically using the first or second order result in $U$ for 
$\Gamma^{(\rm pp)}$, and without introducing the approximation in Eq.~(\ref{eq:o0000}).
The figure also shows the analytical result in Eq.~(\ref{eq:mustarproj}). The second 
order result is clearly larger than the first order result.  For $\mu_c \lesssim 0.5$, 
Eq.~(\ref{eq:mustarproj}) describes the full second order calculation rather well,
while for larger $\mu_c$ there are corrections to the analytic result
which make $\mu_c^*$ still larger compared to Eq.~(\ref{eq:mustarproj}).

It is interesting to discuss the origin of the factor $A_{22}$ in Eq.~(\ref{eq:mustarproj}).
In the present language the term ${\rm log}(D/\omega_{\rm ph})$ in the first order calculation
arises from ``folding in'' large frequency contributions from $U$ in Eq.~(\ref{eq:o0}),
which extend to approximately $\omega\sim D$. The second order term in $U$ is ``folded
in'' in a similar way. However, as shown by Eq.~(\ref{eq:o6}), the contribution from
large frequencies is reduced, leading to the effective band width being reduced by some 
factor for the second order contribution. This reduction factor is surprisingly large (1/10).                      
The origin can be seen in Eq.~(\ref{eq:o7}), where the relevant terms are 
a product of the first and second order contributions in the first term,
containing a factor two, and the first order contribution in the second term.
The prefactor two of the logarithm in the first contribution leads to the factor $A_1^2$,
and thereby a very small factor. 

This shows that the second order terms contribute to retardation effects,
like the first order contribution. However, the reduction is substantially less 
efficient, as described by the small factor $A_{22}$. The band width therefore has 
to be very large to make this contribution to retardation efficient.  

\begin{figure}[!htpb]
\centering
\includegraphics[width=0.3\textwidth,angle=270]{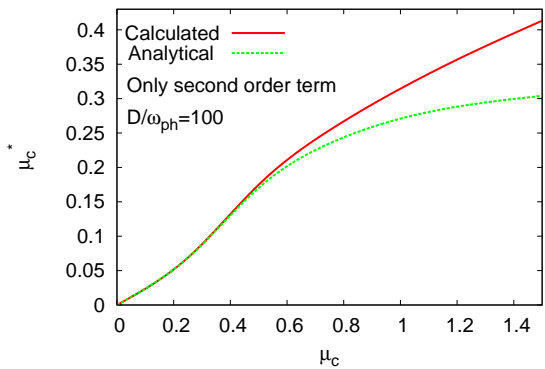}
\caption{(Color online) $\mu^*_c$ as a function of $\mu_c$ for $D/\omega_{\rm ph}=100$ and $\beta \omega_{\rm ph}=240$.
The figure shows the calculated value using just the  second order result for $\Gamma^{(\rm pp)}$ as 
well as the analytical approximation in Eq.~(\ref{eq:o9}).  }
\label{fig:mustarproj2}
\end{figure}

\noindent
It is interesting to consider the second order contribution alone. We can then calculate
results accurate up to order $U^4$. The matrix $\bar A^{-1}$ can be approximated 
by a unit matrix. Making an ansatz for $\mu_c^*$ and identifying with the fourth order
result, we obtain
\begin{equation}
\label{eq:o9}
\mu_c^*=\frac{a\mu_c^2}{1+a\mu_c^2{\rm log}\Big( \frac{DA_2}{\omega_{\rm ph}}\Big)},
\end{equation}
where $A_2=0.20$ was given above. The result of a full calculation is compared with the 
analytical approximation [Eq.~(\ref{eq:o9})] in Fig.~\ref{fig:mustarproj2}. 
Also in this case the analytical result agrees rather well with the full calculation
for $\mu_c \lesssim 0.5$.

\begin{figure}[!htpb]
\centering
\includegraphics[width=0.3\textwidth,angle=270]{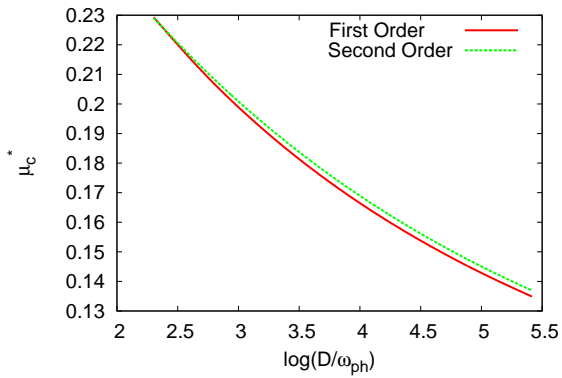}
\caption{(Color online) $\mu_c^*$ as a function of $\log(D/\omega_{\rm ph})$ for $\beta \omega_{\rm ph}=240$
according to a first ($\mu_{c1}=0.5$) and second ($\mu_{c2}=0.2756$) order calculation.
$\mu_{c1}$ and $\mu_{c2}$ were chosen so that the same $\mu_c^*$ was obtained in the two
calculations for $D/\omega_{\rm ph}=10$.}
\label{fig:plotw}
\end{figure}

\noindent
Finally, we compare the first and second order result as a function of $\log(D/\omega_{\rm ph})$
in Fig.~\ref{fig:plotw}. Different values for $\mu_c$ were used in the first and second
order calculations, and chosen such that both calculations gave the same $\mu_c^*$ for 
$D/\omega_{\rm ph}=10$. It is quite interesting that $\mu_c^*$ then decreases in a very similar way
in the two cases.

\noindent
We are now in the position to discuss the results for $\mu_c^*$. We have performed
calculations using the first or second order expression for $\Gamma^{(\rm pp)}$.
We then either calculated Eq.~(\ref{eq:o0}) numerically, without any further approximations.
Alternatively, we performed analytical calculations involving the neglect of additional 
higher order terms in $U$.

When the first order result for $\Gamma^{(\rm pp)}$ is used, the analytical calculations 
can be performed without further approximations. This leads to the Morel-Anderson
result, which has two important features. When $D/\omega_{\rm ph}\to \infty$,                  
$\mu_c^*\to 0$, and when   $U\to \infty$, $\mu_c^*\to 1/{\rm log}(D/\omega_{\rm ph})$ 
stays finite and saturates. Is this still true when the second order contribution to $\Gamma^{(\rm pp)}$
is included?

The analytical results in Eqs.~(\ref{eq:mustarproj},\ref{eq:o9}) have these properties.
However, these results were not derived but are ans\"atze inspired by the Morel-Anderson 
result and adjusted so that they agree with the analytical calculations  to low order in $U$.
Actually, Figs.~{\ref{fig:mustarproj},\ref{fig:mustarproj2} show that although
retardation effects strongly reduce $\mu_c^*$, there is no sign that the values saturate
as $U$ becomes very large. In this sense there is an important difference from the 
Morel-Anderson result. For these large values of $U$, higher order effects in
$\Gamma^{(\rm pp)}$ become important, and it is not clear how these influence the conclusions.
 
The second issue is how $\mu_c^*$ is influenced when $D/\omega_{\rm ph}$ becomes very large.
As is clear from Eqs.~(\ref{eq:mustarproj},\ref{eq:o9}), retardation effects also 
reduce the second order contribution to $\mu_c^*$. This is also seen in Fig.~\ref{fig:plotw}.
However, the effective band width is smaller due to the frequency dependence of the
second order contribution and retardation effects are less efficient.
Nevertheless, the second order contribution goes to zero as $D/\omega_{\rm ph}\to \infty$.
In this context  Fig.~\ref{fig:plotw} may seem surprising. One might have expected 
the second order contribution to drop more slowly with $D/\omega_{\rm ph}$. To understand 
this one can study Eqs.~(\ref{eq:mustarproj},\ref{eq:o9}) and find the $\mu_{c1}$ 
and $\mu_{c2}$ which lead to the same $\mu_c^*$ in the first and second order calculation.
Because of the less efficient retardation effects for the second order term, $\mu_{c2}$
has to be chosen smaller than would otherwise have been the case. The criteria for the
choice of $\mu_{c2}$ is, however, independent of $D$. Thus the two curves in Fig.~\ref{fig:plotw}
should be identical according to Eqs.~(\ref{eq:mustarproj},\ref{eq:o9}). The small 
deviation is due to the inaccuracies of these analytical results for finite $U$.
Fig.~\ref{fig:plotw}, nevertheless, nicely illustrates how both the first and second 
order contributions are systematically reduced as $D/\omega_{\rm ph}$ is increased.

\subsection{Calculations for the gap at $T=0$}
\label{sec:analyticgapT0}
In this section we conduct a complimentary analysis to extract results
for $\mu_c^*$. We calculate the spectral gap of the superconductor
$\Delta_{\rm sp}$ at $T=0$, and show how $\mu_c^*$ enters naturally
in the analytical description. We present this analysis to order
$U^2$. The approach here is similar 
to the original work by Morel and Anderson,\cite{MA62} which included
only the first order term in $U$ and was carried out in the $T=0$
formalism.  Here, we work on the imaginary axis in the limit $T\to 0$. 
Starting point is the self-consistency equation for the off-diagonal
self-energy,   
\begin{equation}
    \Sigma_{21}(i\omega_n)=\frac{1}{\beta}\sum_{m}G_{21}(i\omega_m)
    K(i\omega_n,i\omega_m) ,
\label{eq:selfconSig21}
\end{equation}
where the kernel $K(i\omega_n,i\omega_m)$ includes the following terms,
\begin{equation}
  K(i\omega_n,i\omega_m)=g^2D(i\omega_n-i\omega_m)+U
  -U^2\Pi(i\omega_n+i\omega_m).
\label{eq:kernel}
\end{equation}
The second and the third term are as given in Eq. (\ref{eq:sig21u1})  
and Eq. (\ref{eq:sig21u2bmod}). The expression for $G_{21}(i\omega_m)$
was given in Eq. (\ref{eq:G21}) and a semi-elliptic DOS is used in the
following. We do not consider vertex corrections of the
electron-phonon vertex here, and we simply use the form, 
\begin{equation}
  g^2D(i\omega_n)=
  -\frac{\lambda}{\rho_0}\frac{1}{1+\big(\frac{\omega_n}{\omega_{\rm ph}}\big)^2} .
\end{equation}
The effect of the diagonal self-energy is taken into account in the
analytical calculations for completeness with a $Z$-factor for small
frequencies, $|\omega_n|<\omega_{\rm ph}$. In the numerical calculations in this section it is neglected. 

The self-consistency equation (\ref{eq:selfconSig21}) can be solved numerically
by iteration to find a solution for $\Sigma_{21}(i\omega_n)$. 
For an analytical solution, we need to make some approximations.
At half filling  we use for the Green's function for $|\omega_n|<\omega_{\rm ph}$,
\begin{equation}
  G_{21}(i\omega_n)\simeq -\frac{1}{t}
\frac{\Sigma_{21}(i\omega_n)}{\sqrt{Z^2\omega_n^2+\Sigma_{21}(i\omega_n)^2}}
\end{equation}
for $\omega_{\rm ph}<|\omega_n|<D$,
\begin{equation}
  G_{21}(i\omega_n)\simeq -\frac{1}{t}
\frac{\Sigma_{21}(i\omega_n)}{|\omega_n|}
\end{equation}
and for $|\omega_n|>D$
\begin{equation}
  G_{21}(i\omega_n)\simeq -
\frac{\Sigma_{21}(i\omega_n)}{\omega_n^2} \simeq 0.
\end{equation}
As discussed, $\Pi(i\omega)$ is well approximated by the form given in
Eq. (\ref{pisimpform2}). As noted by Morel and Anderson\cite{MA62} the
$\omega$-dependence of the  off-diagonal self-energy is very similar
to the one of the pairing kernel $K(i\omega_n,i\omega_m)$. Hence, a
suitable ansatz for the off-diagonal self-energy is, 
\begin{equation}
  \Sigma_{21}(i\omega)=\Delta_3+\frac{\Delta_2}{1+b_1|\omega|+b_2\omega^2}+
\frac{\Delta_1-\Delta_2-\Delta_3}{1+\big(\frac{\omega}{\omega_{\rm ph}}\big)^2}.
\end{equation}
It contains the parameters $\Delta_i$. We assume that $b_1$ and $b_2$
take the same values as what is found for  
$\Pi(i\omega)$ in Eq. (\ref{pisimpform2}). By comparing the
$\omega$-dependence with the full numerical solution we find reasonable
accuracy for this assumption. In general we have to solve then for
the parameters three $\Delta_1$, $\Delta_2$, and $\Delta_3$ by evaluating the
self-consistency equation at suitable values of $i\omega$. Unfortunately, the
general case is algebraically very involved. It will be discussed to
some detail in the appendix. Here we only treat the first and purely second
order cases to see the major effects. 

For the first order case, we set $\Delta_2=0$ and omit the
$U^2$-term in Eq.~(\ref{eq:kernel}). We use the self-consistency equations 
$\Sigma_{21}(0)=\Delta_1$, and $\Sigma_{21}(iD) \simeq \Delta_3$. 
When evaluating the $\Sigma_{21}(0)$ and $\Sigma_{21}(iD)$ according to
Eq. (\ref{eq:selfconSig21}), we approximate the integrals in order to find an
analytic solution (see appendix). 
We also assume $\Delta_i\ll\omega_{\rm ph}\ll D$ for simplification.
Then for a self-consistent solution the parameters $\Delta_1$ and
$\Delta_3$ have to satisfy the equations 
\begin{eqnarray}
  \Delta_1&=&
  \frac{\lambda-\mu_c}{Z}\Delta_1\log\Big(\frac{2Z\omega_{\rm
      ph}}{\Delta_1}\Big)+a_0\lambda\Delta_3-\mu_c\Delta_3\log\Big(\frac{D}{\omega_{\rm
      ph}}\Big) ,
\nonumber  \\ 
\Delta_3&=& -\mu_c\Big[\frac{\Delta_1}{Z}\log\Big(\frac{2 Z \omega_{\rm
    ph}}{\Delta_1}\Big)+\Delta_3 \log\Big(\frac{D}{\omega_{\rm
    ph}}\Big)\Big]. 
\label{eq:selfconU}
\end{eqnarray}
The coefficient $a_0$ is given in the appendix.
This yields the non-trivial solutions 
\begin{equation}
  \Delta_1=2 Z\omega_{\rm ph}\e^{-\frac{Z}{\lambda-\mu_c^*(1+a_0\lambda)}},
  \;\;
  \Delta_3=-\frac{2 Z \omega_{\rm ph}\mu_c^*}{\lambda-\mu_c^*}\e^{-\frac{Z}{\lambda-\mu_c^*(1+a_0\lambda)}},
\end{equation}
where the standard result for $\mu_c^*$ is obtained,
\begin{equation}
  \mu_c^*=\frac{\mu_c}{1+ \mu_c\log(\frac{D}{\omega_{\rm ph}})}.
\label{eq:mustarfirstord}
\end{equation}
Note that the gap in the spectral function $\Delta_{\rm sp}$, found from the
pole of the Green's function, occurs at $|\omega|=\Delta_1/Z$ in this approximation.

Accounting for the approximations made in this derivation we can use
this form  with three fitting parameters as in
Eq.~(\ref{eq:ADeq}),\cite{McM68} 
\begin{equation}
  \Delta_{\rm sp} = c_1 \omega_{\rm ph}\e^{-\frac{Z
      c_2}{\lambda-\mu_c^*(1+c_3\lambda)}},
\label{eq:gapanafitform}
\end{equation}
where $c_1>0$, $c_2>1$. We can determine the parameters $c_1$, $c_2$ by
fitting to the numerical solution of Eq. (\ref{eq:selfconSig21}) for $\mu_c=0$ in the regime
$0<\lambda<0.5$. Note that in contrast to the analytical solution the result
of the numerical solution depends in general on the bandwidth $W$. It only becomes
independent in the large bandwidth limit.\cite{Mar90}

For $\mu_c=0$, $D=8/\omega_{\rm ph}=80$, we find with $c_1=1.7$, $c_2=1.07$ good
agreement of the formula (\ref{eq:gapanafitform}) with the numerical results for $\lambda
< 0.5$ as can be seen in Fig. \ref{fig:gap_lambdadep}. Note that for the
Holstein model and larger values of $\lambda$, the result in
Eq. (\ref{eq:gapanafitform}) 
underestimates $\Delta_{\rm sp}$ as already pointed out by Allen and
Dynes.\cite{AD75} 

\begin{figure}[!htpb]
\centering
\includegraphics[width=0.45\textwidth]{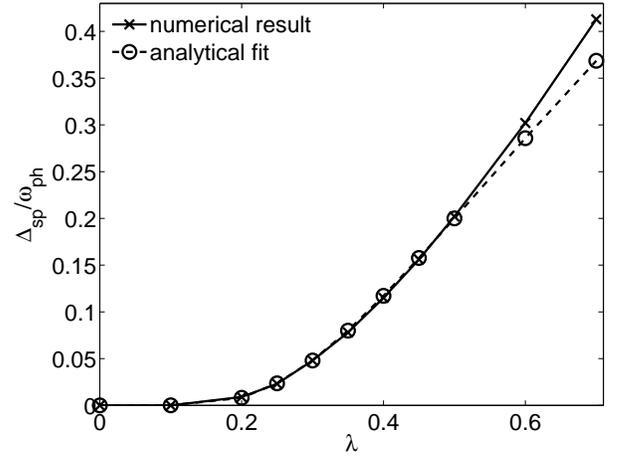}
\caption{The spectral gap $\Delta_{\rm sp}$ as calculated from the numerical solution
  of Eq. (\ref{eq:selfconSig21}) as a function of $\lambda$ for
  $D/\omega_{\rm  ph}=80$ in comparison with the analytical form in
  Eq. (\ref{eq:gapanafitform}) with the the parameters $c_1=1.7$,
  $c_1=1.07$.}        
\label{fig:gap_lambdadep}
\end{figure}
\noindent
For fixed $\lambda=0.5$, we also compare the $\mu_c$-dependence of the analytical result in
Eq. (\ref{eq:gapanafitform}) with the full numerical solution. In
Fig. \ref{fig:gapmucdepUU2onlyU} a comparison can be found, where we used the
same values for $c_1$ and $c_2$ as for $\mu_c=0$ and found that $c_3=0.8$
gives a reasonable fit.

\begin{figure}[!htpb]
\centering
\includegraphics[width=0.45\textwidth]{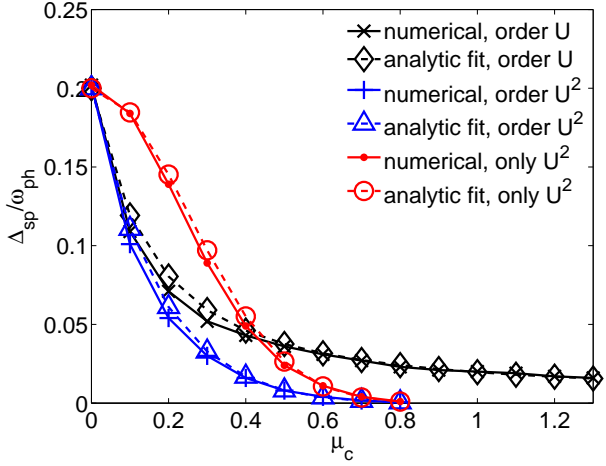}
\caption{(Color online) The spectral gap $\Delta_{\rm sp}$ as calculated from the numerical solution
  of Eq.~(\ref{eq:selfconSig21}) with different kernels as a  function
  of $\mu_c$ for $D/\omega_{\rm  ph}=80$ in comparison with the
  analytical result in Eq. (\ref{eq:gapanafitform}).}        
\label{fig:gapmucdepUU2onlyU}
\end{figure}
\noindent
We analyze the situation including only the $U^2$-term now. We set
$\Delta_3=0$ in this case and omit the constant $U$-term in Eq.~(\ref{eq:kernel}). 
To determine the parameters $\Delta_1$ and $\Delta_2$, we use the
following two conditions: $\Sigma_{21}(0)=\Delta_1=\sum_i N_{1i}\Delta_i$, 
and $\Sigma_{21}(iD)+\Sigma_{21}(-iD) \simeq 2\Delta_2/(1+b_1D+b_2D^2)=\sum_i N_{2i}\Delta_i$. 
This yields the self-consistency equation,
\begin{equation}
  1=N_{11}+\frac{N_{12}N_{21}}{\frac{2}{1+b_1D+b_2D^2}-N_{22}}.
\label{eq:NU2eq}
\end{equation}
Making similar approximation as in the first order calculation we find
for the coefficients $N_{ij}$, 
\begin{equation}
  N_{11}=\frac1{Z}(\lambda-a\mu_c^2)\log(T_1),
\end{equation}
and
\begin{equation}
  N_{21}=-\frac{2}{Z}\frac{a\mu_c^2}{1+b_1D+b_2D^2}\log(T_1).
\end{equation}
with 
\begin{equation}
  T_1=\frac{Z\omega_{\rm ph}+\sqrt{(Z\omega_{\rm ph})^2+\Delta_1^2}}{\Delta_1}.
\label{eq:T1}
\end{equation}
One also finds 
\begin{equation}
  N_{12}=a_{12}\lambda-a\mu_c^2\log\Big(\frac{D}{\omega_{\rm ph}}\bar A_{12}\Big),
\end{equation}
and 
\begin{equation}
  N_{22}=-a\mu_c^2 \frac{2}{1+b_1D+b_2D^2}
 \log\Big(\frac{D}{\omega_{\rm ph}}\bar A_{22}\Big).
\end{equation}
The expressions for $a_{12}$, $\bar A_{12}$ and for $\bar A_{22}$ can
be found in the appendix. The coefficients $\bar A_{12}<1$ and $\bar
A_{22}<1$ appear due to additional terms to $1/\omega$ in the
integrand which accelerate the decay towards higher energy. Hence,
they lead to a reduced effective bandwidth, $D\to\bar A_{22} D $ as
discussed in the previous section. 
The solution of Eq. (\ref{eq:NU2eq}) yields
\begin{equation}
  \Delta_1=2 Z\omega_{\rm
    ph}\e^{-\frac{Z}{\lambda-\mu_c^*(1+a_{12}\lambda)-\mu_{c,1}^*}},
\label{eq:resD1U2}
\end{equation}
with
\begin{equation}
  \mu_c^*=\frac{a\mu_c^2 }{1+a\mu_c^2 \log\Big(\frac{D}{\omega_{\rm
        ph}}\bar A_{22}\Big)},
\end{equation}
and
\begin{equation}
  \mu_{c,1}^*=\frac{\log(\frac{\bar A_{22}}{\bar
      A_{12}})a^2\mu_c^4}{1+a\mu_c^2 \log\Big(\frac{D}{\omega_{\rm
        ph}}\bar A_{22}\Big)}.
\label{eq:mustarho}
\end{equation}
The result for $\mu_c^*$ has the same form as
what was derived in Eq. (\ref{eq:o9}). In addition a higher order term
$\mu_{c,1}^*$ appears. In the first order calculations all higher
order contributions cancel in the numerator of the expression for
$\mu_c^*$. However, in the second order calculation this is not the
case anymore and an additional term remains. The coefficient $\log(\bar
A_{22}/\bar A_{12})$ does not increase with the bandwidth $W$ and therefore
the whole term  becomes small in the large bandwidth limit due to the
logarithm in the denominator. For the relevant values one has $\bar A_{22}/\bar
A_{12}\approx 2$ such that the coefficient of the $\mu_c^2$ is
relatively small and the term does not contribute much for small
values of $\mu_c$. For larger values, 
however, it does play a role. Thus it can account for the discrepancy
between the numerical and the analytical result from Eq. (\ref{eq:o9}) observed in
Fig. \ref{fig:mustarproj2}, where the numerical result does not seem
to saturate.

We would like to check these analytical findings with the full
numerical solution of the self-consistency equation. Formally, we had
found very similar results to the ones of the last section derived
from the projection scheme. For the comparison with the numerics we
take these expressions and use the values for the parameters derived
there, i.e. we use $\bar A_{22}\to A_2$ , where $A_{2}=0.197$ or
$\log(A_{22})=-1.624$. We omit the term $\mu_{c,1}^*$, and for
$c_1$, $c_2$ we take the same values as above, and we use
$a_{12}=c_3$. The reason for this procedure is that due to a number of
approximations involved in the analytic calculation the results for
these coefficient do not tend to be very accurate. Moreover, we aim
for a unified description with as little parameters as possible.

The result for the ``only $U^2$''-calculation is added in
Fig. \ref{fig:gapmucdepUU2onlyU} and compared with the full numerical
solution. We find good agreement. For small values of $\mu_c$ the reduction of
$\Delta_{\rm sp}$ is smaller than for the first order term, but then $\Delta_{\rm sp}$
drops more rapidly. Notice that the second order result for
$\mu_c^*$ is analogous in the form to the first order calculation. The 
difference is the factor $\bar A_{22}<1$ in the logarithm in the
denominator. Hence, the retardation effects are less effective in this
case as discussed in the previous section. The results here are
derived independently from the arguments of the last
section, but are fully consistent with them. The higher order term
$\mu_{c,1}^*$ is not very important for the values of
$\mu_c$ appearing in Fig. \ref{fig:gapmucdepUU2onlyU}. 

In the general case including both the first and second order terms in
Eq.~(\ref{eq:kernel}), we have to solve for three parameters and hence
have three self-consistency equations to be solved. In general this
can be written as a matrix equation $\vct \Delta=\underline M \vct
\Delta$.  Algebraically this becomes rather lengthy and yields a
number of different terms for $\mu_c^*$, as discussed in the appendix. 
To simplify the discussion, we use results in the form of
Eq.~(\ref{eq:gapanafitform}) with $\mu_c^*$ as introduced in
Eq.~(\ref{eq:mustarproj}), 
\begin{equation}
  \mu_c^*=\frac{\mu_c+a\mu_c^2 }{1+\mu_c\log(\frac{D}{\omega_{\rm ph}})+a\mu_c^2 \log\Big(\frac{D}{\omega_{\rm ph}}A_{22}\Big)},
\end{equation}
for comparison with the numerical results.
In Fig. \ref{fig:gapmucdepUU2onlyU} we have included the numerical
result of Eq. (\ref{eq:selfconSig21}) with the full kernel in
Eq. (\ref{eq:kernel}). We also included the  analytical description  
based on Eq. (\ref{eq:gapanafitform}) and (\ref{eq:mustarproj}) with
the same value $\log(A_{22})=-2.28$ or $A_{22}=0.1022$ as in the
previous section. We find a rather good agreement for the range of
values of $\mu_c$. 

We can accurately calculate the dependence of the spectral gap
$\Delta_{\rm sp}$ on $\mu_c$  for given $\lambda$, $D$ and
$\omega_{\rm ph}$ numerically. However, it is not possible to
calculate $\mu_c^*$ directly from the self-consistency equation
(\ref{eq:selfconSig21}). If we assume that the form
of Eq.~(\ref{eq:gapanafitform}), which neglects higher 
order terms of the form Eq.~(\ref{eq:mustarho}), gives a good
description and we can solve for $\mu_c^*$,  
\begin{equation}
\mu_c^*=\frac{\lambda}{1+c_3 \lambda}+\frac{Zc_2}{\log\Big(\frac{\Delta_{\rm
      sp}}{c_1\omega_{\rm ph}}\Big)(1+c_3\lambda)}.
\label{eq:mustargapanafitform}
\end{equation}
As higher order terms are neglected this is not the complete result
for $\mu_c^*$ for the whole range of $\mu_c$. However, if $\mu_c^*$ is
interpreted as the quantity in competition with $\lambda$ to cause
superconductivity, then this form is useful and the inversion of
Eq. (\ref{eq:gapanafitform}) can give us an estimate for $\mu_c^*$. 

The results for $\mu_c^*$ obtained from Eq. (\ref{eq:mustargapanafitform})
together with the analytical estimates in
Eqs. (\ref{eq:mustar},\ref{eq:o9},\ref{eq:mustarproj}) can be found in
Fig. \ref{fig:mustar_mucdepUU2onlyU2}. 

\begin{figure}[!htpb]
\centering
\includegraphics[width=0.45\textwidth]{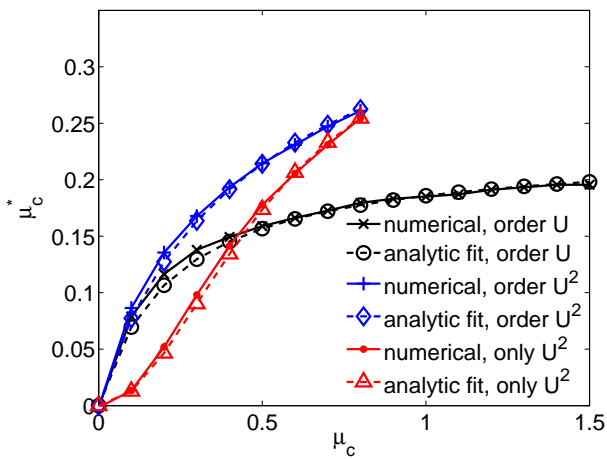}
\caption{(Color online) $\mu_c^*$ as a function of $\mu_c$ for
  $D/\omega_{\rm ph}=80$ computed from
  Eq. (\ref{eq:mustargapanafitform}). We   show in comparison the
  result for the first order and the  second order calculation. The
  fit parameters are $c_1$, $c_2$ and $c_3$ are the same as above.}       
\label{fig:mustar_mucdepUU2onlyU2}
\end{figure}
\noindent
The results look very similar to the ones in the previous section in
Fig. \ref{fig:mustarproj}.
For small $\mu_c$, $\mu_c^*$ increases linearly with $\mu_c$ in
Eqs. (\ref{eq:mustar},\ref{eq:mustarproj}). This implies an 
initially relatively large drop of superconductivity once the Coulomb
repulsion becomes finite (see Fig. \ref{fig:gapmucdepUU2onlyU}). However,
then the curves bend and the analytical results saturate. Only for
very small values of $\mu_c$ the first and 
second order results are in agreement, otherwise the second order result is
substantially larger. For the first order
calculation the upper boundary is given by $\mu_{c, \rm
  max}^*=1/\log(\frac{D}{\omega_{\rm ph}})$ independent of $\mu_c$. It is
reached both in the limit of large $\mu_c$ and large ratio $D/\omega_{\rm
  ph}$. For instance for $D/\omega_{\rm  ph}=1000$, we have $\mu_{c, \rm
  max}^*\approx 0.145$, similar to what is usually used in the
literature. Using $\mu_c=0.5$ for the screened Coulomb interaction gives with
the same ratio for $D/\omega_{\rm ph}$ the value $\mu_c^*=0.112$ (see
Tab. \ref{table:mustar1}). 

\begin{table}[!htpb]
\begin{tabular}{| l| c| c| c |}\hline
  $D/\omega_{\rm ph}$ & 10 & 100 & 1000  \\ \hline
 \; $\mu_c^*$ in Eq. (\ref{eq:mustar}) \;& \; 0.232 \;&\; 0.151 \;&\; 0.112 \; \\ \hline
 \; $\bar\mu_c^*$ in Eq. (\ref{eq:mustarbar}) \;& \; 0.286 \; &\; 0.172 \; &  \; 0.123\; \\ \hline
 \; $\mu_c^*$ in Eq. (\ref{eq:mustarproj}) \;& \; 0.386 \; &\; 0.204 \; &  \; 0.139\; \\ \hline
\end{tabular}
 \caption{Exemplary values for $\mu_c^*$ for $\mu_c=0.5$.}
  \label{table:mustar1}
\begin{tabular}{| l| c| c| c |}\hline
  $D/\omega_{\rm ph}$ & 10 & 100 & 1000  \\ \hline
 \; $\mu_c^*$ in Eq. (\ref{eq:mustar}) \;& \; 0.303 \;&\; 0.172 \;&\; 0.127 \; \\ \hline
 \; $\bar\mu_c^*$ in Eq. (\ref{eq:mustarbar}) \;& \; 0.366 \; &\; 0.199 \; &  \; 0.136\; \\ \hline
 \; $\mu_c^*$ in Eq. (\ref{eq:mustarproj}) \;& \; 0.700 \; &\; 0.268 \; &  \; 0.166 \; \\ \hline
\end{tabular}
 \caption{Exemplary values for $\mu_c^*$ for $\mu_c=1$.}
  \label{table:mustar2}
\end{table}

We can see that the higher order results for $\mu_c^*$ are generally
substantially larger than in the first order calculation. In particular the
results are larger than the simple minded estimate, $\bar \mu_c^*$ in
Eq. (\ref{eq:mustarbar}), where retardation effects are the same for the first
and second order term. 
For this expression the limits of large bandwidth and large $\mu_c$ lead to the same
result $\bar\mu_{c, \rm  max}^*=1/\log(\frac{D}{\omega_{\rm ph}})$, which is however reached
already for smaller values, e.g. for  $\mu_c=0.5$ we have $\bar\mu_c^*=0.1233$. 
In contrast, the result in Eq. (\ref{eq:mustarproj}) goes to $\mu_{c,
  \rm  max}^*=1/\log(\frac{D}{\omega_{\rm ph}}A_{22})$ in the 
limit of large $\mu_c$ and to $\mu_{c, \rm  max}^*=1/\log(\frac{D}{\omega_{\rm ph}})$ in the
limit of a very large bandwidth. With the estimate for $A_{22}$ above we find for 
$D/\omega_{\rm  ph}=1000$  that in the large $\mu_c$ limit $\mu_{c, \rm
  max}^* \approx 0.216$ is about 50 \% larger than the first order
estimate. For comparison we give some results in tables
\ref{table:mustar1} and \ref{table:mustar2} for 
$\mu_c=0.5$, $\mu_c=1$, and the ratios $D/\omega_{\rm ph}=10,100,1000$.
Notice, that the more accurate result for $\mu_c^*$ as
obtained from the numerical calculation and shown in
Fig. \ref{fig:mustarproj} can be still significantly larger than the
estimate in Eq. (\ref{eq:mustarproj}). We conclude that the usual
result in Eq. (\ref{eq:mustar}) substantially underestimates $\mu_c^*$
for intermediate and larger values of $\mu_c$. However, for very large values
of $D/\omega_{\rm ph}$ retardation effects are operative in all cases and lead
to a strongly reduced value of $\mu_c^*$.

\subsection{Calculations for $T_c$}
For completeness we also include a brief section on the critical
temperature $T_c$.  It is analyzed similar to the last section.
The basis for the calculations is the pairing matrix in
Eq. (\ref{symmatrixinst}). For the pairing vertex we use the same terms as in
the previous section in Eq. (\ref{eq:kernel}),
\begin{equation}
   \Gamma^{(\rm pp)}(i\omega_{n_1},i\omega_{n_2};0)= -K(i\omega_{n_1},i\omega_{n_2}).
\end{equation}
The effect of the $Z$-factor is neglected. Then with the bare Green's function
and a semi-elliptic DOS 
\begin{equation}
  \tilde\chi^0(i\omega_{n_1};0,0)=-\frac{G(i\omega_{n_1})}{i\omega_{n_1}}=
\rho_0\frac{\pi}{2t}\left(\sqrt{1+\frac{4t^2}{\omega_n^2}}-1\right).
\end{equation}
$\Pi(i\omega_{n})$ is calculated numerically from the free Green's function. With this we
compute the matrix $M_{n_1,n_2}$ in Eq. (\ref{symmatrixinst}) and search for
the largest eigenvalue. We compare the results for $T_c$ for three different
calculations: (1) including only the $U$-term, (2) including only the
$U^2$-term and (3) including both. 

Apart from the first order calculation it is not easy to find a good
analytic approximation for the eigenvalue equation (\ref{eveq}). In
principle, one can do something similar to what has been done in the last
section and make an appropriate ansatz for the eigenvector. To first order
in $U$ this works reasonably well. We find a result of the standard form, 
\begin{equation}
  T_c=c_1\omega_{\rm ph}\e^{-\frac{Zc_2}{\lambda-\mu_c^*(1+c_3\lambda )}}.
\label{eq:Tcfitform}
\end{equation}
where $\mu_c^*$ is given by Eq. (\ref{eq:mustarfirstord}). 
For the higher order analysis we did not pursue an analytical solution, and
instead also assume $T_c$ as in Eq. (\ref{eq:Tcfitform}) and for $\mu_c^*$ the
form in Eq. (\ref{eq:mustarproj}).

First we fix the constants $c_1$ and $c_2$ by fitting to the result for
$\mu_c=0$, see Fig. \ref{fig:Tc_lambdadep}. We take the value
$c_1=1/1.2\approx 0.833$ as in Ref. \onlinecite{AD75} and
find a good fit for $c_2=1.04$. Similar as before the agreement is only good
up to values of $\lambda<0.5$.

\begin{figure}[!htpb]
\centering
\includegraphics[width=0.45\textwidth]{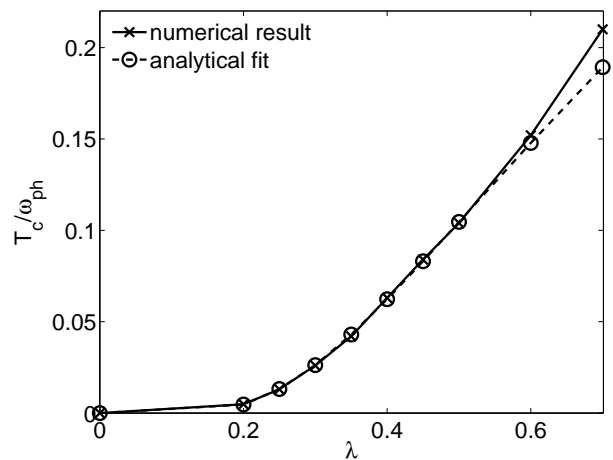}
\caption{(Color online) $T_c$ as a function of $\lambda$ for $D/\omega_{\rm
    ph}=80$. }       
\label{fig:Tc_lambdadep}
\end{figure}
\noindent
In Fig. \ref{fig:Tc_mucdepUU2onlyUdep} we give the numerical
result for $T_c$ as a function of $\mu_c$ for the first order calculation,
only the second order and first plus second order. We have kept
$\lambda=0.5$ constant.

\begin{figure}[!htpb]
\centering
\includegraphics[width=0.45\textwidth]{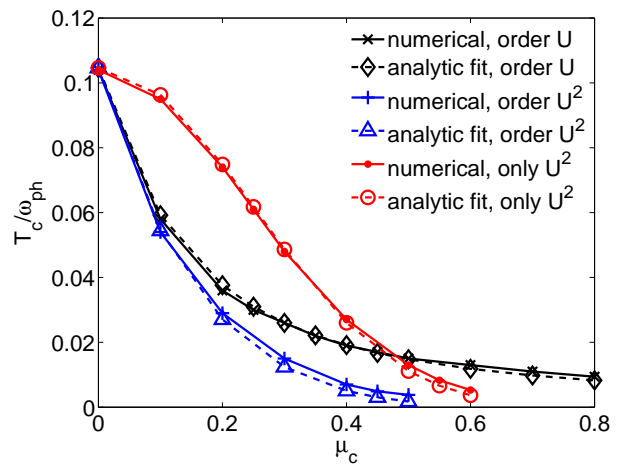}
\caption{(Color online) $T_c$ as a function of $\mu_c$ for $D/\omega_{\rm
    ph}=80$.}       
\label{fig:Tc_mucdepUU2onlyUdep}
\end{figure}
\noindent
$T_c$ decays in a very similar way as the spectral gap when $\mu_c$ is increased.
We included results from the analytic fit form in Eq. (\ref{eq:Tcfitform})
with the value $c_3=0.8$. For the first order calculation,
we find good agreement with the numerical result for the case
$\lambda=0.5$. 
For the second order results we use the same parameters as before and
the fits are reasonable, i.e. they show that the functional form is
close to the actual result. 

Note, that in spite of the identical form for the pairing kernel, the
results for $T_c$ here and the ones for the gap at $T=0$ in the
previous section are obtained from two independent calculations. The
results indicate that for the range of parameters studied the analytical
expressions Eqs. (\ref{eq:mustar},\ref{eq:o9},\ref{eq:mustarproj})
describe the effect of $\mu_c^*$ on $T_c$ quite accurately. In 
particular the results are very similar to what has been found in the
projective approach before, and hence a consistent picture emerges. We
conclude that higher order dynamic effects, and in particular 
the second order contributions, give an important correction to the usual
results for the Coulomb pseudopotential. Not only does the higher
order term give a direct increase of the coupling, as would be the
case in expression in Eq. (\ref{eq:mustarbar}) it also leads to a
reduced effective bandwidth in the logarithm in the denominator. This
is an effect which to our knowledge has not been discussed in the
literature so far. In the following section we analyze how these
effects are manifest in the non-perturbative  DMFT calculations. 

\section{DMFT and perturbative results}

In this section we put the analysis of the previous sections together and
compare with non-perturbative DMFT calculations, which include all possible
renormalization effects. We first clarify that it is therefore 
very important to work with renormalized parameters for the interpretation of
the results. Then we demonstrate up to which interaction strengths the
perturbative results derived in the previous section are reliable.  

We focus on calculations for the spectral gap $\Delta_{\rm sp}$ at $T=0$ and
consider a half filled band. The spectral gap $\Delta_{\rm sp}$ is extracted
directly from the gap edges of the diagonal spectral function. It is usually
well approximated by the product $z\Sigma_{21}(0)$, but it can be a bit larger
due to the  frequency dependence of the off-diagonal self-energy.
For an interpretation of the DMFT results we need to compare with the 
perturbation theory (PT) results.    
We include the following terms for the self-consistent PT calculation: For the
diagonal and off-diagonal self-energy we use Eqs. (\ref{eq:sig11withvertex}) and 
(\ref{eq:sig21withvertex}). The vertex is approximated by the contributions
from RPA screening in Eq.~(\ref{eq:gamscrrpa}) up to second order in $U$ and
the second order in $U$ corrections in Fig.~\ref{correctelphgU2}.  
For a small phonon frequency, the $\omega=0$ value of the vertex function
provides a good approximation.  
From the diagrams in $U$ we take Eqs. (\ref{eq:sig21u1}), and
(\ref{eq:sig11u2a})-(\ref{eq:sig21u2b}) into
account. The phonon propagator is taken as an input from DMFT calculations and
not calculated self-consistently, similar to what has been done in
Ref.~\onlinecite{BHG11}.

To get a feeling for the renormalization effects let us first
consider calculations,  where the bare parameters
$\lambda_0=\rho_02g^2/\omega_0$ and $\omega_0$ in the Hamiltonian
(\ref{hubholham}) are kept fixed. For $\mu_c=0$ the system has a
superconducting solution and we analyze how the gap $\Delta_{\rm sp}$ 
is affected, when $\mu_c$ becomes finite.
For $\lambda_0=0.308$ and $\lambda_0=0.382$, the corresponding spectral gaps
$\Delta_{\rm  sp}$ are shown in Fig. \ref{fig:gapconslambda0_mucdep}. 
For comparison we have also included the results from the self-consistent PT
as explained above in Fig.~\ref{fig:gapconslambda0_mucdep}. These are seen to
be in very good agreement with the DMFT result.

\begin{figure}[!htpb]
\centering
\includegraphics[width=0.44\textwidth]{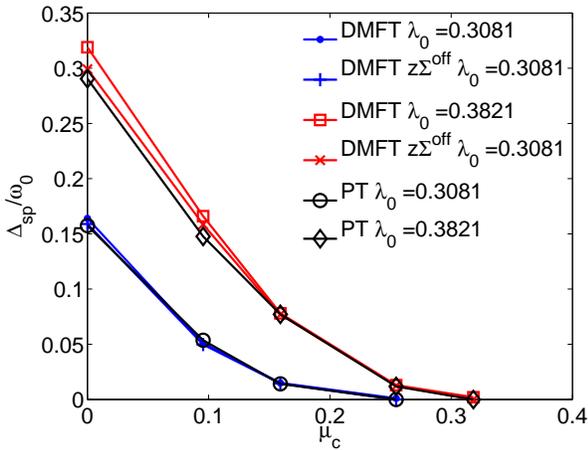}
\caption{(Color online) Behavior of the spectral gap $\Delta_{\rm sp}$ and
  $z\Sigma^{\rm off}(0)$ for constant $\lambda_0=0.308$ and $\lambda_0=0.382$
  ($\omega_0=0.1$ in both cases) as a function of $\mu_c$.}       
\label{fig:gapconslambda0_mucdep}
\end{figure}
\noindent
We find a rapid decrease of $\Delta_{\rm sp}$ when $\mu_c$ becomes finite,
similar to the results in Fig.~\ref{fig:gapmucdepUU2onlyU}. However, 
notice that $\Delta_{\rm sp}$ goes to zero for even smaller values of
$\mu_c \sim 0.3$. Naively one could conclude that $\mu_c^*$ is even
larger than what was discussed in the last section even though we are
still at very weak coupling in $U$. As we will see, however, this drastic
reduction of $\Delta_{\rm sp}$ has a different origin. It can be understood
by analyzing the relevant effective parameters and the PT results.    
 
Since, as shown in Sec.~\ref{sec:resmustar} many aspects of the PT are well described
by approximate analytical results, such as Eq. (\ref{eq:mustarproj}) for $\mu_c^*$ or
Eq. (\ref{eq:gapanafitform}) for $\Delta_{\rm sp}$, it makes sense to use
those equations to analyze the results. In Sec.~\ref{sec:resmustar} the
equations were used in terms of bare parameters, but to 
understand the DMFT results we need to use renormalized parameters. 
Those effective parameters extracted from the DMFT calculation are shown in
Fig.~\ref{fig:renpar_conslambd0_mucdep} as function of $\mu_c$. 

\begin{figure}[!htpb]
\centering
\subfigure[]{\includegraphics[width=0.23\textwidth]{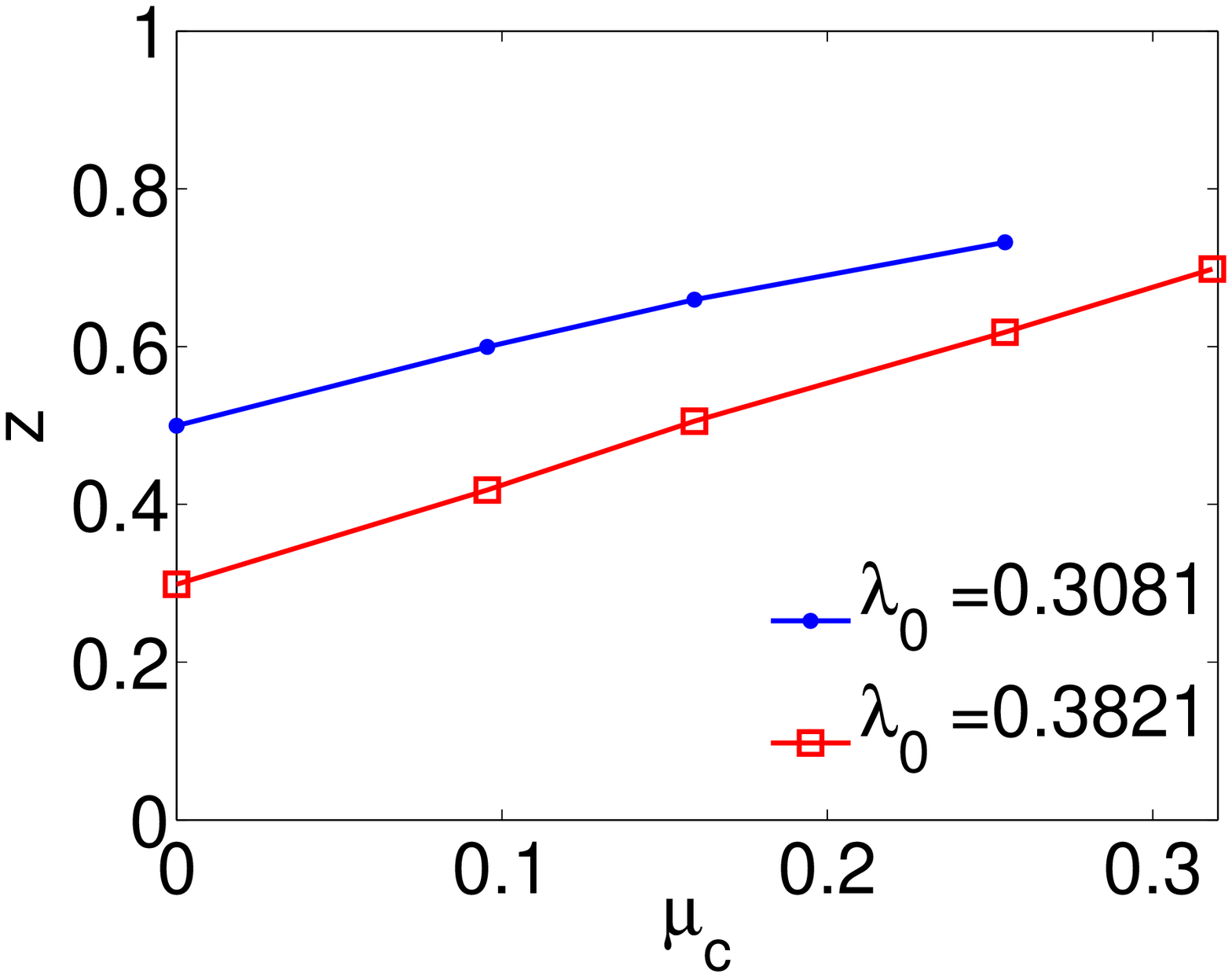}}
\subfigure[]{\includegraphics[width=0.23\textwidth]{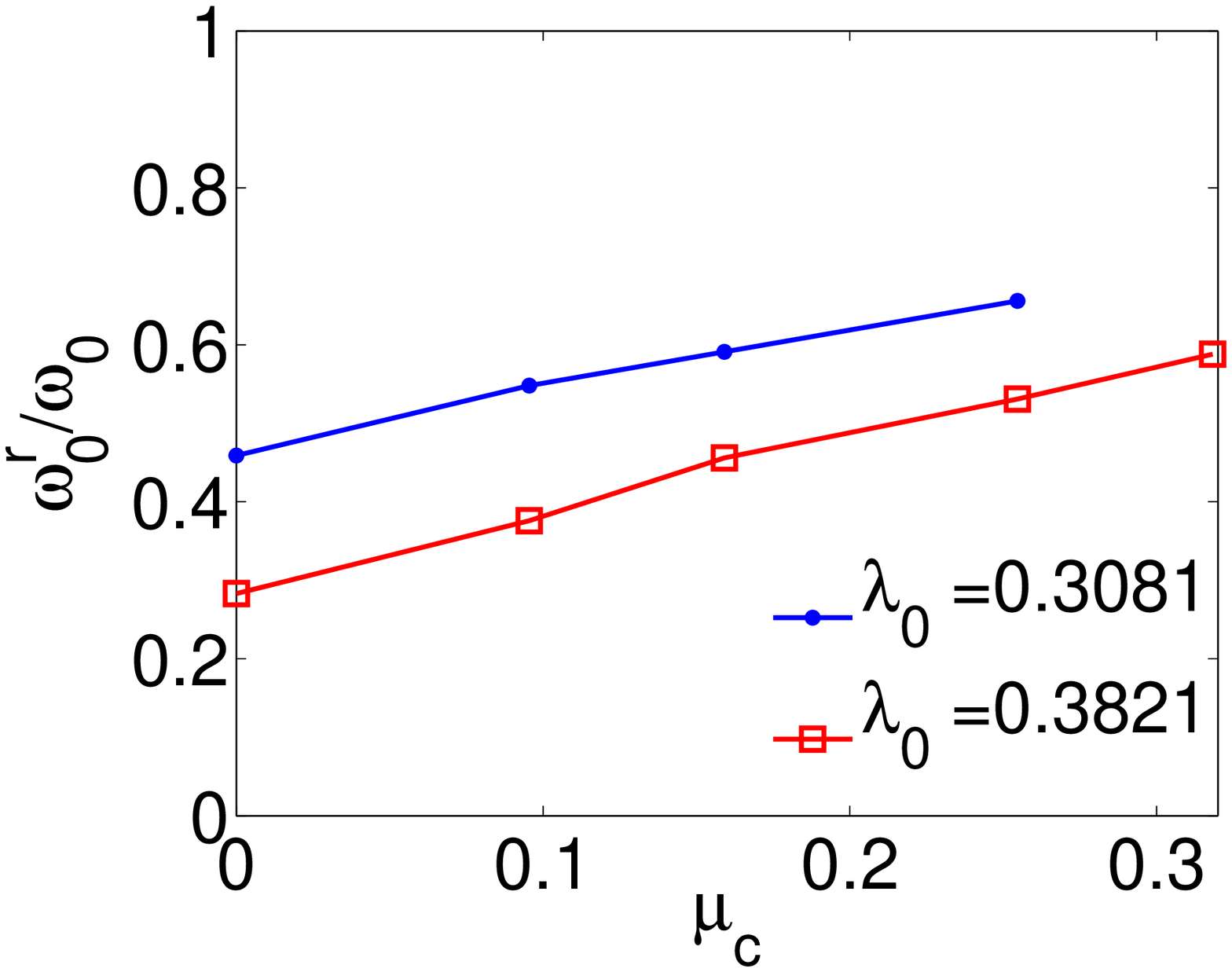}}
\subfigure[]{\includegraphics[width=0.23\textwidth]{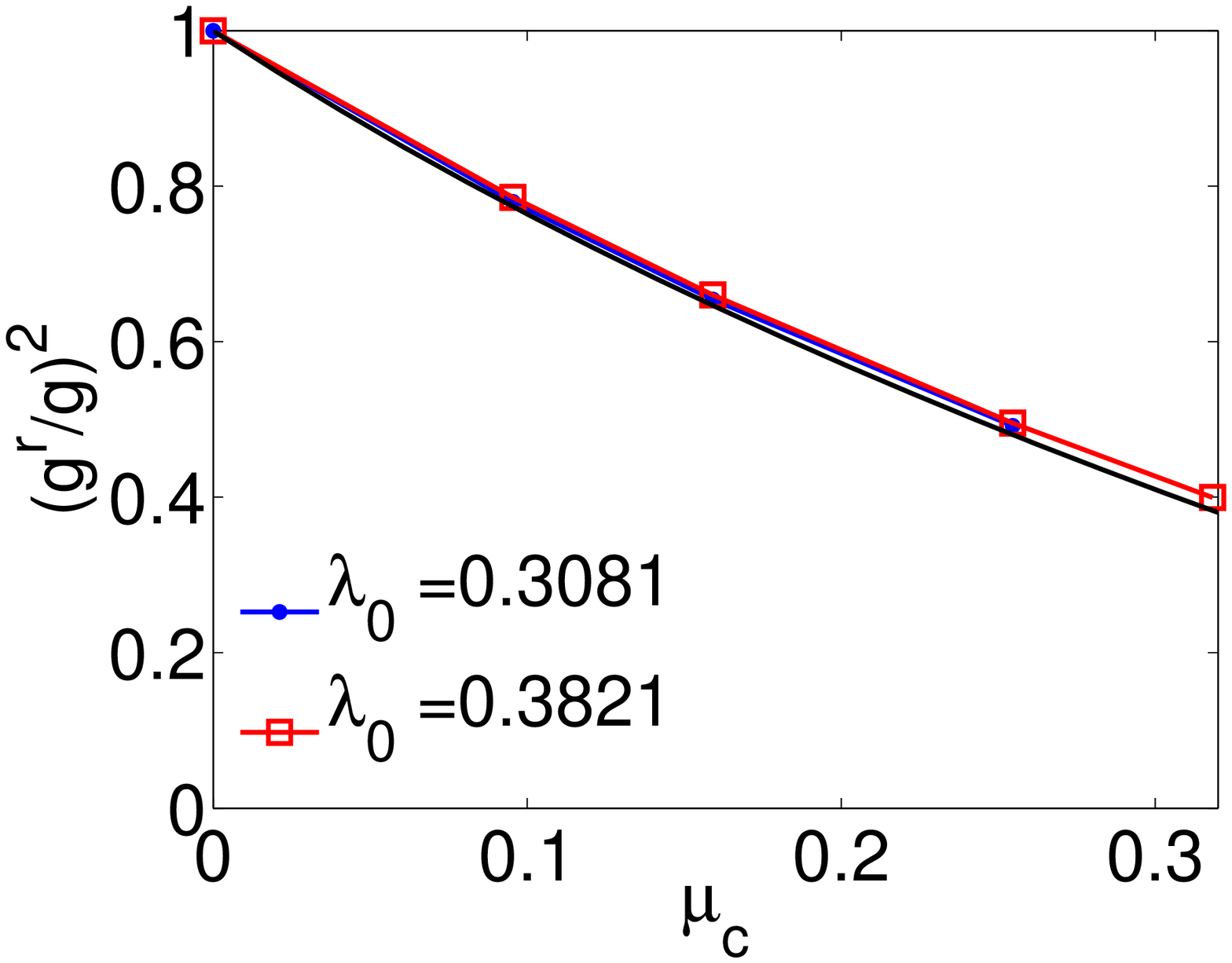}}
\subfigure[]{\includegraphics[width=0.23\textwidth]{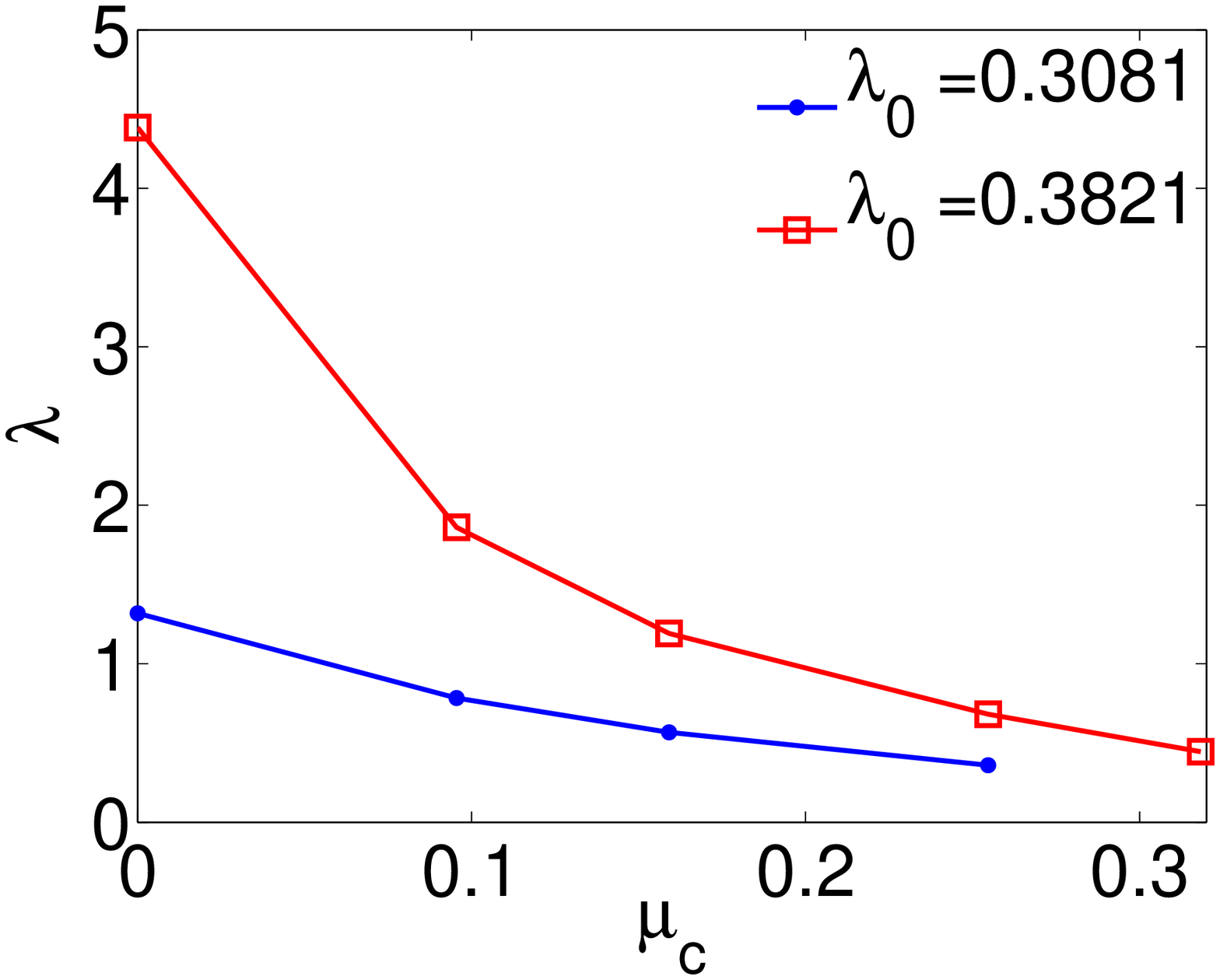}}
\caption{(Color online) The effective parameters (a) the quasiparticle renormalization $z$,
  (b) the renormalized phonon frequency $\omega_0^r/\omega_0$  (c) the
  renormalized coupling constant $(g^r/g)^2$ (the full line corresponds to
  the result with $a$ and $b$ obtained from the free Green's functions) and (c) the
  effective $\lambda$  as a function of $\mu_c$.}       
\label{fig:renpar_conslambd0_mucdep}
\end{figure}
\noindent
The $z$-factor in Fig.~\ref{fig:renpar_conslambd0_mucdep} (a) increases
moderately with $\mu_c$. According to Eq. (\ref{eq:gapanafitform}) with
$Z=z^{-1}$ this would actually help superconductivity, so it can not be
responsible for the reduction in Fig.~\ref{fig:gapconslambda0_mucdep}. The
renormalized phonon frequency $\omega_0^r$ 
in Fig.~\ref{fig:renpar_conslambd0_mucdep} (b) also increases with   
$\mu_c$ showing that corrections to the phonon self-energy and electron-phonon
vertex from finite $U$ are important. In the prefactor of
Eq. (\ref{eq:gapanafitform}) with $\omega_{\rm ph}=\omega_0^r$ this leads to
an enhancement of $\Delta_{\rm sp}$, whereas $\mu_c^*$ increases with
$\omega_0^r$ which leads to a reduction of $\Delta_{\rm sp}$. Both effects are
not strong enough to be decisive. 
The strongest and most important effect of $\mu_c$ can be seen in
Fig.~\ref{fig:renpar_conslambd0_mucdep} (c) and (d), where we plot the
renormalized coupling $g^r$ from Eq. (\ref{eq:effvertcorr}) 
and the effective $\lambda$ according to Eq. (\ref{lambdafinU}). The
parameters $a$ and $b$ in Eq. (\ref{eq:effvertcorr}) are 
calculated from the full Green's functions. We find that the renormalized coupling
$g^r$ decreases substantially with $\mu_c$. In addition the phonon self-energy
is modified for finite $U$. Therefore, the effective $\lambda$ becomes
much smaller. So the main reason for the rapid suppression of $\Delta_{\rm
  sp}$ is the strong effect of the corrections to the electron-phonon
vertex and to the phonon self-energy, such that the effective $\lambda$
decreases.

Using values $\omega_{\rm ph}=\omega_0^r$ we can calculate results
for the Coulomb pseudopotential $\mu_c^*$ according to
Eq. (\ref{eq:mustarproj}). 
The ratio of electron over phonon scale is with $D/\omega_0^r\sim 35$ not as
large as in the last section, and thus retardation effects not as effective.
For the largest value $\mu_c\simeq 0.318$ we find $\mu_c^*\simeq
0.2$. At this $\mu_c$ we have $\lambda\simeq 0.45$ and the analytic
expression in Eq. (\ref{eq:gapanafitform}) yields $\Delta_{\rm
  sp}/\omega_0\simeq 0.0002$, where $Z=z^{-1}$, $\omega_{\rm
  ph}=\omega_0^r$, and for the fitting parameters $c_i$ take the same values
as in Sec.~\ref{sec:analyticgapT0}. This is in agreement 
with the DMFT finding that superconductivity goes to zero then. If we
compare the results for the spectral gap according to
Eq.~(\ref{eq:gapanafitform}) with the renormalized parameters in
Fig.~\ref{fig:renpar_conslambd0_mucdep}  with the results from the full
calculation in Fig.~\ref{fig:gapconslambda0_mucdep} we find good
agreement. Hence we conclude that the superconducting state at small $\mu_c$
can be well understood in terms of the effective parameters $\lambda$,
$\omega_0^r$, $z$ and $\mu_c^*$ and the approximate equations for $\mu_c^*$
and $\Delta_{\rm  sp}$.  

The conclusion up to this stage is that the strongest effect of the Coulomb
repulsion in the DMFT calculations is to renormalize the effective $\lambda$
via electron-phonon vertex and phonon propagator such that superconductivity
drops to zero rapidly. Even at very weak  
coupling these effects play an important role in the Hubbard-Holstein model and
must be taken into account.  Since in this work our main interest is the
effectiveness of 
retardation effects visible in the direct competition between
$\lambda$ and $\mu_c^*$, we will offset the vertex-correction
effect in the following. We do this by appropriately adjusting the
bare parameters, i.e., increasing $\lambda_0$ and $\omega_0$. We proceed by
keeping $\lambda=\lambda(\mu_c)$ as defined in Eq. (\ref{lambdafinU}) constant. 
For $g^r$ we rely on the perturbative results. This was found to give a
relatively good description up to $U\sim W/2$.\cite{HHAS03}  First we
do a calculation for certain bare values $\omega_0$ and $\lambda_0$ at
$\mu_c=0$. From the phonon spectral function we can then extract 
the value $\omega_0^r$ and $\lambda$ from (\ref{lambdafinU}) where,
$g^r=g$. Then we choose a finite value of $\mu_c$. We have to increase $\lambda_0$
and $\omega_0$, such that $\omega_0^r$ roughly equals the $U=0$ value
and $\lambda$ remains approximately the same according to
Eq. (\ref{lambdafinU}). For the renormalized coupling $g^r$ entering
Eq. (\ref{lambdafinU}) we consider two conditions: (a) the second order
expansion as in Eq. (\ref{eq:effvertcorr2}) and (b) the RPA series
plus the second order terms as in Eq. (\ref{eq:effvertcorr}).
With the set of bare parameters $(\lambda_0,\omega_0,\mu_c)$ found
from this procedure we also do perturbative calculations for
comparison.  We distinguish PT-a, which includes up to second order
diagrams for the electron-phonon vertex, and PT-b which includes the
RPA series plus the second order terms for the electron-phonon
vertex. The results of such a calculation, where $\lambda\simeq 1$
according to condition (a) and $\omega_0^r/D\simeq 0.025$ are shown in 
Fig.~\ref{fig:gapconslambdar_mucdep} as $\Delta_{\rm sp}$ vs $\mu_c$.    

\begin{figure}[!htpb]
\centering
\includegraphics[width=0.44\textwidth]{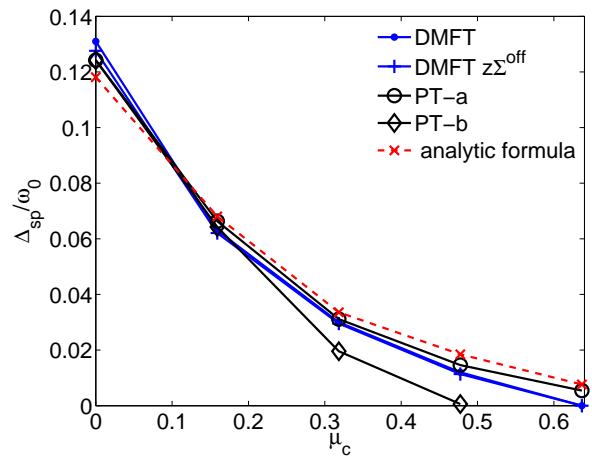}
\caption{(Color online) Behavior of the DMFT result for the spectral
  gap $\Delta_{\rm sp}$ and $z\Sigma^{\rm off}(0)$ for constant
  $\lambda\simeq 1$ according to condition (a) for the vertex
  corrections (see text),   $\omega_0^r\simeq 0.05$ as a function of
  $\mu_c$ in comparison with PT-a and PT-b (see text) and the analytic
  formula in Eq. (\ref{eq:gapanafitform}).}       
\label{fig:gapconslambdar_mucdep}
\end{figure}
\noindent
The DMFT results show a steady decrease of $\Delta_{\rm sp}$ on increasing
$\mu_c$. Since the effective $\lambda$ and $\omega_{\rm ph}$ are kept
constant the diminution of $\Delta_{\rm sp}$ is now due to the
competition with the Coulomb repulsion $\mu_c$. To understand the
result quantitatively we compare it with the perturbative calculations
PT-a, PT-b and the analytic results based on
Eq. (\ref{eq:gapanafitform}) using the effective parameters and $\mu_c^*$ according to
Eq. (\ref{eq:mustarproj}) with the value of $A_{22}$ as in
Sec.~\ref{sec:resmustar}. We find a 
relatively good agreement of the DMFT result with PT-a and the
analytic formula up to $\mu_c\sim 0.5$. This demonstrates that (i) the
effective parameter description is appropriate, (ii) the
electron-phonon vertex correction according to condition (a) is
suitable and (iii) that the derived higher order form for the Coulomb
pseudopotential in Eq. (\ref{eq:mustarproj}) captures correctly the
results of the PT and the of 
the full DMFT calculation.  This validates the analysis of the
Sec.~\ref{sec:resmustar} in a more complete calculation and it
corroborates our  findings for $\mu_c^*$ and retardation effects by
comparison with non-perturbative DMFT up to intermediate values of
$\mu_c\sim 0.5$.  For larger values of $\mu_c$, higher order correction 
enhance the value of $\mu_c^*$ such that $\Delta_{\rm sp}$ is suppressed
stronger. In addition the electron-phonon vertex is not well described by
condition (a) anymore. 
The PT-b calculation agrees with DMFT quite well up to values
$\mu_c\sim 0.3$, but then overestimates the reduction of the
electron-phonon vertex and therefore leads to a too strong suppression
of superconductivity.

To extend the present analysis to larger values of
$\mu_c$, one needs to find reliable estimates for the full renormalized
electron-phonon vertex $\Gamma^{(\rm ep)}_U$. A higher order perturbative analysis or
non-perturbative calculation for this quantity could provide this.
A consistent calculation up to certain order in $\mu_c$ should then also
include higher order corrections to the self-energies, which will complicate
the analytical calculation for $\mu_c^*$ further. This is beyond the scope of
this work.  

\section{Conclusions}
For the occurrence of conventional superconductivity it is important that on
the one hand there is a sizeable electron-phonon coupling and on the other
that the detrimental effects of the Coulomb repulsion are reduced sufficiently
by screening and retardation effects. We have stressed in this article that
while the effect of the former can be described in a controlled fashion by
Migdal-Eliashberg theory, the standard approach to the latter is based on an
uncontrolled approximation, since there is no Migdal theorem for the Coulomb
interaction.   
For the Hubbard-Holstein model we have analyzed this issue in a controlled
framework by a combination of perturbative calculations and non-perturbative
DMFT calculations. 
We have shown that the conventional arguments based on the
lowest order diagrams  become modified when higher order corrections are taken into
account. There is still a reduction of the Coulomb repulsion due to 
logarithmic terms in the denominator. This demonstrates that indeed a small
value of $\mu_c^*$ can be obtained due to retardation effects even when higher
order corrections are considered. Thus our results support the 
arguments by Morel and Anderson qualitatively. However, the effective energy scale 
separation is reduced. This is due to the dynamic behavior of the
higher order diagrams which make retardation effects less
operative. This result was shown explicitely in two independent
calculations in a combination of analytical and numerical
arguments. By studying the occurrence of superconductivity with DMFT, where
all higher order corrections are present, we were able corroborate our
findings up to intermediate coupling strength. The perturbative approach
allowed us to distinguish different renormalization effects. Our analysis
is limited to intermediate coupling strength due to the difficulty
to reliably estimate the vertex corrections to the electron-phonon
vertex. Non-perturbative calculations for this quantity would be
desirable. 

We conclude that the usual expression for $\mu^*_c$ in
Eq.~(\ref{eq:mustar}) is only valid for  small values for $\mu_c$. The
second order corrections lead to less efficient retardation effects as
shown in Eq. (\ref{eq:mustarproj}),
\begin{equation}
  \mu_c^*=\frac{\mu_c+a\mu_c^2 }{1+\mu_c \log\Big(\frac{E_{\rm el}}{\omega_{\rm
        ph}}\Big)+a\mu_c^2 \log\Big(\alpha\frac{E_{\rm el}}{\omega_{\rm ph}}\Big)}, 
\end{equation}
where we found $\alpha\approx 0.1$ for a typical large energy separation of
electron and phonon scale $E_{\rm el}/\omega_{\rm
  ph}$. The coefficient $a$ is given by the limit
$\omega\to0$ of the particle-hole bubble $\Pi$ divided by $\rho_0$, the DOS at
the Fermi energy, and $\alpha$ can be estimated from the decay
of  $\Pi$ with  $\omega$. In addition higher order terms
appear such that $\mu_c^*$ does not saturate in the limit of large
$\mu_c$. Thus, for systems with sizeable effective Coulomb repulsion
we should expect larger values for $\mu^*_c$ than the traditional quote
$\mu_c^*\sim 0.1$. As a consequence the values for  $\mu^*_c$ are not as
universal as sometimes claimed and predictions for $T_c$ can be unreliable.

It is premature to draw detailed conclusion from our calculations for real
materials such as lithium, since the Hubbard model is not an accurate
description for itinerant metallic systems, which are better described by
an electron gas model. Nevertheless one can understand our results as
a qualitative trend, which shows that for system with less efficient screening,
such that $\mu_c$ is larger, we expect an enhanced value of
$\mu_c^*$ as compared to  what is traditionally quoted.
This could be expected in systems with larger values of $r_s$. Hence our conclusions
would seem to be line with the observation in a number of
materials, such as Li with enhanced $\mu_c^*$, and what is quoted in work
based on DFT calculations.\cite{SS96} 
Similar conclusions are important for systems where the ratio $E_{\rm
  el}/\omega_{\rm   ph}$ is reduced from the usual scenario, which could be
the case in picene.  
We would like to stress, however, that a more accurate attempt of
understanding the problem should also include the dynamic effect of screening
the bare Coulomb repulsion in a metal, which was not taken into account
explicitely.  As discussed in the introduction this can lead to very small
and even negative values of $\mu_c^*$. We expect that the combination of
these effects and the corrections studied here, which lead to an enhancement,
will eventually lead to the physical values occurring in nature. More detailed
calculations are required to fully resolve this quantitatively.   
As long as $\mu_c^*$ can not be estimated reliably, the predictive power of
the theory of electron-phonon superconductivity is limited.

\bigskip
\noindent{\bf Acknowledgment}\par
\noindent
We wish to thank N. Dupuis, A.C. Hewson, P. Horsch, C. Husemann, G. Kotliar,
D. Manske, F. Marsiglio, A.J. Millis, G. Sangiovanni, and  R. Zeyher
for helpful discussions. JB acknowledges financial
support from the DFG through BA 4371/1-1, and JH acknowledges
support from the grant NSF DMR-0907150.

\begin{appendix}
\section{Details for the calculation of the spectral gap}
In this appendix we collect some details for the analytic calculation
in Sec.~\ref{sec:analyticgapT0}.
\subsection{Integrals for the first order case}
For $\Sigma_{21}(0)$ we approximate the integrals over small frequencies,
\begin{equation}
  \frac{1}{2\pi}\integral{\omega}{-\omega_{\rm ph}}{\omega_{\rm ph}}
G_{21}(i\omega)g^2D(-i\omega) \simeq 
\frac{\Delta_1\lambda}{Z}\log (T_1)\; ,
\label{eq:ap1}
\end{equation}
and
\begin{equation}
  \frac{1}{2\pi}\integral{\omega}{-\omega_{\rm ph}}{\omega_{\rm ph}}
G_{21}(i\omega)U = \frac{\Delta_1\mu_c}{Z}\log (T_1)\; ,
\label{eq:ap2}
\end{equation}
where $T_1$ is given in (\ref{eq:T1}). For larger frequencies we write
\begin{equation}
  \frac{1}{2\pi}\Big[\integral{\omega}{-D}{-\omega_{\rm ph}}
+\integral{\omega}{\omega_{\rm ph}}{D}\Big]
G_{21}(i\omega)g^2D(-i\omega)
=a_0 \Delta_3\lambda \; ,
\label{eq:ap3}
\end{equation}
and
\begin{equation}
  \frac{1}{2\pi}\Big[\integral{\omega}{-D}{-\omega_{\rm ph}}
+\integral{\omega}{\omega_{\rm ph}}{D}\Big]
G_{21}(i\omega)U
= -\Delta_3\mu_c \log\Big(\frac{D}{\omega_{\rm ph}}\Big) \; .
\label{eq:ap4}
\end{equation}
We have
\begin{equation}
  a_0= \integral{\omega}{\omega_{\rm ph}}{D}
\frac{1}{\omega}\frac{1}{1+\frac{\omega^2}{\omega_{\rm
      ph}^2}}=\frac{1}{2}\log\Big(2\frac{D^2}{\omega_{\rm
    ph}^2+D^2}\Big).
\label{eq:a0}
\end{equation}
For $\Sigma_{21}(iD)$, the electron-phonon contribution is small for
$\omega_{\rm ph}/D\ll 1$, and the Coulomb contribution is the same as
above. This yields Eq. (\ref{eq:selfconU}).

\subsection{Integrals for the only second order case}
We give some results for the second order calculation. 
To determine the coefficient $\bar A_{12}$ we calculate
\begin{equation}
  \integral{\omega}{\omega_{\rm ph}}{D}
\frac{1}{\omega}\frac{1}{f_1(\omega)^2}=\log\Big(\frac{D}{\omega_{\rm   ph}}\bar A_{12}\Big),
\label{eq:ap8}
\end{equation}
where
\begin{equation}
  f_1(\omega)=1+b_1|\omega|+b_2\omega^2.
\end{equation}
One finds that
\begin{eqnarray}
 \bar A_{12}&=&\sqrt{\frac{f_1(\omega_{\rm ph})}{f_1(D)}} 
 \exp\Big[\frac{1}{B_1^2}\Big(\frac{2b_2-(b_1^2+b_1b_2D)}{f_1(D)}-
 \nonumber \\
&& \frac{2b_2-(b_1^2+b_1b_2\omega_{\rm
     ph})}{f_1(\omega_{\rm  ph})}\Big)\Big]
\exp\Big(\frac{b_1^3-6b_1b_2}{B_1^2}f_a\Big) \nonumber
 \end{eqnarray}
with $B_1=\sqrt{4b_2-b_1^2}$  and
\begin{equation}
  f_a=\arctan\Big(\frac{b_1+2b_2D}{B_1}\Big)-\arctan\Big(\frac{b_1+2b_2\omega_{\rm
      ph} }{B_1}\Big). 
\end{equation}
By comparing the integrals in Eq. (\ref{eq:ap4}) and
Eq. (\ref{eq:ap8}) we see that the factor $0<\bar A_{12}<1$
leads to a reduction of the effective bandwidth. For the given fitting
values $b_1$, $b_2$, and large ratios $D/\omega_{\rm ph}>100$ one
finds $\bar A_{12}\approx 0.2$, which can be compared to the
calculation in Eq. (\ref{eq:o8}).  
The coefficient $\bar A_{22}$ is obtained from the integral
\begin{eqnarray*}
&&  \integral{\omega}{\omega_{\rm ph}}{D}
\frac{1}{\omega}\frac{1}{f_1(\omega)}
\Big(\frac{1}{f_1(D+\omega)}+\frac{1}{f_1(D-\omega)}\Big) \\
&&=\frac{2\log(\frac{D}{\omega_{\rm  ph}}\bar A_{22})}{f_1(D)}.
\end{eqnarray*}
The integral on the left hand side can be carried out analytically,
but the expression is lengthy and not instructive. We can express
$\bar A_{22}$ as
\begin{eqnarray*}
\bar A_{22}=\exp\Big[ \integral{\omega}{\omega_{\rm ph}}{D}
\frac{1}{\omega}\Big(\frac{f_1(D)}{2f_1(\omega)}
\Big(\frac{1}{f_1(D+\omega)}+\frac{1}{f_1(D-\omega)}\Big)-1\Big)\Big].
\end{eqnarray*}
Since the function $1/f_1(D+\omega)+1/f_1(D-\omega)$ increases in the integration
interval the coefficient $\bar A_{22}$ comes out larger than $\bar
A_{12}$. For the parameters above we find $\bar A_{22}\approx 0.44$.
The coefficient for $\lambda$ reads
\begin{equation}
   a_{12}= \integral{\omega}{\omega_{\rm ph}}{D}
\frac{1}{\omega}\frac{1}{1+\frac{\omega^2}{\omega_{\rm ph}^2}}\frac{1}{f_1(\omega)}
\end{equation}
The integral can be solved analytically but the expression is
lengthy. Due to the reduction factor $1/f_1(\omega)$, $a_{12}$ term is 
a bit smaller than $a_0$ in Eq. (\ref{eq:a0}).

\subsection{Calculation up to second order}
We use the following three conditions: (i) $\Sigma_{21}(0)=\Delta_1$,
(ii),
\begin{equation}
 \Sigma_{21}(i\bar b_1)+\Sigma_{21}(-i \bar b_1)\simeq
2\Delta_3+2\Delta_2/f_1(\bar b_1),  
\end{equation}
with $\bar b_1\equiv1/b_1 $,
and (iii),
\begin{equation}
\Sigma_{21}(iD)+\Sigma_{21}(-iD)\simeq 2\Delta_3+2\Delta_2/f_1(D).     
\end{equation}
We use $\Sigma_{21}(0)=\sum_i N_{1i}\Delta_i$,
\begin{eqnarray}
&\Sigma_{21}(i\bar b_1)+\Sigma_{21}(-i \bar b_1)=\sum_i N_{2i}\Delta_i,  \\
&\Sigma_{21}(iD)+\Sigma_{21}(-iD)=\sum_i N_{3i}\Delta_i.  
\end{eqnarray}
This implies $N_{1i}=M_{1i}$, 
\begin{eqnarray}
M_{21}=f_1(\bar b_1)N_{21}/2, M_{22}=f_1(\bar
b_1)N_{22}/2  , \\
M_{23}=f_1(\bar b_1)(N_{23}-2)/2,  
M_{31}=N_{31}/2,\\
M_{32}=(N_{32}-2/f_1(D))/2, M_{33}=N_{33}/2.
\end{eqnarray}
The calculations give with certain approximations in the integrals
\begin{equation}
  N_{11}=\frac1{Z}(\lambda-\mu_c-a\mu_c^2)\log(T_1),
\end{equation}
\begin{equation}
  N_{12}=a_{12}\lambda-\mu_c\log\Big(\frac{D}{\omega_{\rm ph}}A_{12}^{(1)}\Big)-a\mu_c^2\log\Big(\frac{D}{\omega_{\rm ph}}A_{12}^{(2)}\Big),
\end{equation}
\begin{equation}
  N_{13}=\frac{1}{2}\lambda-\mu_c\log\Big(\frac{D}{\omega_{\rm ph}}\Big)-a\mu_c^2\log\Big(\frac{D}{\omega_{\rm ph}}A_{13}^{(2)}\Big),
\end{equation}
\begin{equation}
  N_{21}=\frac2{Z}\Big(\lambda\frac{\omega_{\rm ph}^2}{\bar b_1^2}-\mu_c-\frac{a\mu_c^2}{f_1(\bar b_1)}\Big)\log(T_1), 
\end{equation}
\begin{equation}
  N_{22}=a_{22}\lambda-2\mu_c\log\Big(\frac{D}{\omega_{\rm
      ph}}A_{22}^{(1)}\Big)-\frac{2a\mu_c^2}{f_1(\bar b_1)}\log\Big(\frac{D}{\omega_{\rm ph}}A_{22}^{(2)}\Big),
\end{equation}
\begin{equation}
  N_{23}=a_{23}\lambda-2\mu_c\log\Big(\frac{D}{\omega_{\rm
      ph}}\Big)-\frac{2a\mu_c^2}{f_1(\bar
    b_1)}\log\Big(\frac{D}{\omega_{\rm ph}}A_{23}^{(2)}\Big), 
\end{equation}
\begin{equation}
  N_{31}=-\frac2{Z}\Big(\mu_c+\frac{a\mu_c^2}{f_1(D)}\Big)\log(T_1),
\end{equation}
\begin{equation}
  N_{32}=a_{32}\lambda-2\mu_c\log\Big(\frac{D}{\omega_{\rm
      ph}}A_{32}^{(1)}\Big)-\frac{2a\mu_c^2}{f_1(D)}\log\Big(\frac{D}{\omega_{\rm ph}}A_{32}^{(2)}\Big),
\end{equation}
\begin{equation}
  N_{33}=a_{33}\lambda-2\mu_c\log\Big(\frac{D}{\omega_{\rm
      ph}}\Big)-\frac{2a\mu_c^2}{f_1(D)}\log\Big(\frac{D}{\omega_{\rm
      ph}}A_{33}^{(2)}\Big).  
\end{equation}
The coefficients $a_{12}$, $A_{12}^{(2)}=\bar A_{12}$ and
$A_{32}^{(2)}=\bar A_{22}$ were given above. The others read,
\begin{equation}
  A_{12}^{(1)}=A_{32}^{(1)}=A_{22}^{(1)}=A_{13}^{(2)}=\sqrt{\frac{f_1(\omega_{\rm ph})}{f_1(D)}}\exp\Big[\frac{b_1}{B_1}f_a\Big],
\end{equation}
\begin{equation}
   a_{22}=\integral{\omega}{\omega_{\rm ph}}{D}
\frac{1}{\omega}\frac{1}{f_1(\omega)}\Big(\frac{1}{1+\frac{(\bar
    b_1+\omega)^2}{\omega_{\rm ph}^2}}+\frac{1}{1+\frac{(\bar
    b_1-\omega)^2}{\omega_{\rm ph}^2}} \Big),
\end{equation}
\begin{equation}
   a_{23}=\integral{\omega}{\omega_{\rm ph}}{D}
\frac{1}{\omega}\Big(\frac{1}{1+\frac{(\bar
    b_1+\omega)^2}{\omega_{\rm ph}^2}}+\frac{1}{1+\frac{(\bar
    b_1-\omega)^2}{\omega_{\rm ph}^2}} \Big),
\end{equation}
\begin{equation}
   a_{32}=\integral{\omega}{\omega_{\rm ph}}{D}
\frac{1}{\omega}\frac{1}{f_1(\omega)}\Big(\frac{1}{1+\frac{(D+\omega)^2}{\omega_{\rm
      ph}^2}}+\frac{1}{1+\frac{(D-\omega)^2}{\omega_{\rm ph}^2}}\Big),
\end{equation}
\begin{equation}
   a_{33}=\integral{\omega}{\omega_{\rm ph}}{D}
\frac{1}{\omega}\Big(\frac{1}{1+\frac{(D+\omega)^2}{\omega_{\rm
      ph}^2}}+\frac{1}{1+\frac{(D-\omega)^2}{\omega_{\rm ph}^2}}\Big), 
\end{equation}
\begin{equation}
A^{(2)}_{22}=\exp\Big[ \integral{\omega}{\omega_{\rm ph}}{D}
\frac{1}{\omega}\Big(\frac{f_1(\bar b_1)}{2f_1(\omega)}
\Big(\frac{1}{f_1(\bar b_1+\omega)}+\frac{1}{f_1(\bar b_1-\omega)}\Big)-1\Big)\Big],
\end{equation}
\begin{equation}
A^{(2)}_{23}=\exp\Big[ \integral{\omega}{\omega_{\rm ph}}{D}
\frac{1}{\omega}\Big(\frac{f_1(\bar b_1)}{2}
\Big(\frac{1}{f_1(\bar b_1+\omega)}+\frac{1}{f_1(\bar b_1-\omega)}\Big)-1\Big)\Big],
\end{equation}
\begin{equation}
A^{(2)}_{33}=\exp\Big[ \integral{\omega}{\omega_{\rm ph}}{D}
\frac{1}{\omega}\Big(\frac{f_1(D)}{2}
\Big(\frac{1}{f_1(D+\omega)}+\frac{1}{f_1(D-\omega)}\Big)-1\Big)\Big].
\end{equation}
One can give analytic expressions for the integrals, which are however, lengthy and not
very instructive. For typical values for $D/\omega_{\rm ph}$,
we have $A^{(2)}_{23}\approx 1.6$ and $A^{(2)}_{33}\approx 1.7$. All other
coefficients obey  $0<A^{(\alpha)}_{ij}<1$, leading to a reduce effective
bandwidth as discussed before. The coefficients $a_{ij}$ are small for
$D/\omega_{\rm ph}\gg 1$. 

From the matrix equation $\vct \Delta=\underline M \vct \Delta$, we
can derive the self-consistency equation,
\begin{widetext}
\begin{equation}
  1=M_{11}+\frac{M_{13}M_{31}}{1-M_{33}}+\Big(M_{12}+\frac{M_{32}}{1-M_{33}}\Big)
\frac{M_{21}(1-M_{33})+M_{23}M_{31}}{(1-M_{22})(1-M_{33})-M_{23}M_{32}},
\label{Meq}
\end{equation}
or when using the expression for $M_{11}$ with $T_1=\frac{Z\omega_{\rm
    ph}+\sqrt{(Z\omega_{\rm ph})^2+\Delta_1^2}}{\Delta_1}$ as in Eq. (\ref{eq:T1}),
\begin{equation}
  1=\frac{\log(T_1)}{Z}\Big[\lambda-\mu_c-a\mu_c^2+\frac{M_{13}M_{31}\frac{Z}{\log(T_1)}}{1-M_{33}}
+\Big(M_{12}+\frac{M_{32}}{1-M_{33}}\Big)\frac{Z}{\log(T_1)}  
\frac{M_{21}(1-M_{33})+M_{23}M_{31}}{(1-M_{22})(1-M_{33})-M_{23}M_{32}}\Big].
\label{Meq2}
\end{equation}
\end{widetext}
When solved for gap parameter $\Delta_1$, the term in the square brackets is the
exponent, such that terms to the right of $\lambda$ contribute to the
expression for the pseudopotential $\mu_c^*$. 
We now would like to argue that in an expansion in $\mu_c$ the dominant term
is of the form of Eq.~(\ref{eq:mustarproj}).
We neglect the terms involving $\lambda$ in $M_{ij}$ to
simplify the arguments. We find then that $-\mu_c-a\mu_c^2$ together
with $\frac{M_{13}M_{31}\frac{Z}{\log(T_1)}}{1-M_{33}}$ gives a term
of the form in Eq. (\ref{eq:mustarproj}),
\begin{equation}
  \mu_c^*=\frac{\mu_c+a\mu_c^2 }{1+\mu_c \log\Big(\frac{D}{\omega_{\rm
        ph}}\Big)+a\mu_c^2 \log\Big(\frac{D}{\omega_{\rm ph}}\bar A_{33}^{(2)}\Big)}, 
\label{eq:mustarapp} 
\end{equation}
where
\begin{equation}
\bar A^{(2)}_{33}=\exp\Big[ \integral{\omega}{\omega_{\rm ph}}{D}
\frac{1}{\omega}\Big(\frac{1}{2}
\Big(\frac{1}{f_1(D+\omega)}+\frac{1}{f_1(D-\omega)}\Big)-1\Big)\Big].
\end{equation}
There are also additional terms to order $\mu_c^3$ and $\mu_c^4$
whose coefficients are not proportional to $\log(D/\omega_{\rm ph})$
in the numerator (cf. discussion in Sec.~\ref{sec:analyticgapT0}). The
denominator in the other part, $F_2\equiv(1-M_{22})(1-M_{33})-M_{23}M_{32}$, has
similar properties to the one in Eq.~(\ref{eq:mustarapp}) but contains also
contributions to order $\mu_c^3$ and $\mu_c^4$ and terms $\sim
\log(D/\omega_{\rm ph})^2$. Due to a cancellation the lowest order 
term in the numerator, $M_{21}(1-M_{33})+M_{23}M_{31}$, is $\sim
a\mu_c^2(f_1(\bar b_1)/f_1(D)-1)$. From the prefactor, 
\begin{equation}
\frac{M_{12}(1-M_{33})+M_{32}}{1-M_{33}}  
\end{equation}
the lowest order term is $-1/[f_1(D)(1-M_{33})]$. This gives a contribution
\begin{equation}
 \sim \frac{a\mu_c^2\left(\frac{f_1(\bar b_1)}{f_1(D)}-1\right)}{f_1(D)F_2},
\end{equation}
which is smaller compared to
$a\mu_c^2$. All other terms are of the order $\mu_c^3$ and higher. Hence, in an
expansion in $\mu_c$ the term to the right in the square brackets in
Eq.~(\ref{eq:mustarapp}) gives a smaller contribution to $\mu_c^*$. This
explains why the numerical results in Sec.~\ref{sec:analyticgapT0} were fit well by an
expression involving $\mu_c^*$ of the form in Eq.~(\ref{eq:mustarproj}).

\end{appendix}


\bibliography{artikel,biblio1}

\end{document}